# Two Complementary Methods for Relative Quantification of Ligand Binding Site Burial Depth in Proteins: The 'Cutting Plane' and 'Tangent Sphere' Methods

**Vicente M. Reyes, Ph.D.***
**E-mail:** vmrsbi.RIT.biology@gmail.com

*work done at:

Dept. of Pharmacology, School of Medicine,
University of California, San Diego
9500 Gilman Drive, La Jolla, CA 92093-0636
&
Dept. of Biological Sciences, School of Life Sciences
Rochester Institute of Technology
One Lomb Memorial Drive, Rochester, NY 14623

**Abstract.** We describe two complementary methods to quantify the degree of burial of ligand and/or ligand binding site (LBS) in a protein-ligand complex, namely, the 'cutting plane' (CP) and the 'tangent sphere' (TS) methods. To construct the CP and TS, two centroids are required: the protein molecular centroid (global centroid, GC), and the LBS centroid (local centroid, LC). The CP is defined as the plane passing through the LBS centroid (LC) and normal to the line passing through the LC and the protein molecular centroid (GC). The "exterior side" of the CP is the side opposite GC. The TS is defined as the sphere with center at GC and tangent to the CP at LC. The percentage of protein atoms (a.) inside the TS, and (b.) on the exterior side of the CP, are two complementary measures of ligand or LBS burial depth since the latter is directly proportional to (b.) and inversely proportional to (a.). We tested the CP and TS methods using a test set of 67 well characterized protein-ligand structures (Laskowski et al., 1996), as well as the theoretical case of an artificial protein in the form of a cubic lattice grid of points in the overall shape of a sphere and in which LBS of any depth can be specified. Results from both the CP and TS methods agree very well with data reported by Laskowski et al., and results from the theoretical case further confirm that that both methods are suitable measures of ligand or LBS burial. Prior to this study, there were no such numerical measures of LBS burial available, and hence no way to directly and objectively compare LBS depths in different proteins. LBS burial depth is an important parameter as it is usually directly related to the amount of conformational change a protein undergoes upon ligand binding, and ability to quantify it could allow meaningful comparison of protein dynamics and flexibility.

**Abbreviations:** CP, cutting plane; TS, tangent sphere; CPM, cutting plane method; TSM, tangent sphere method; CPi, CP or CPM index; TSi, TS or TSM index; LBS, ligand binding site; GC, global or molecular centroid; LC, local or LBS centroid; lig, ligand; res, residue; sdc, sidechain; PPI, protein-protein-interaction/s; PLI, protein-ligand interaction/s; PLC, protein-ligand complex

**Keywords:** protein-ligand interactions; ligand burial; ligand binding site burial; protein centroid; cutting plane method; tangent sphere method; protein flexibility; protein dynamics

## 1 Introduction

This work was largely inspired by that of A. Ben-Shimon and M. Eisenstein in 2005. Depth of ligand burial in its cognate receptor protein is an important property of the latter biomolecule because it provides an indication of protein dynamics: deeply buried ligand binding sites imply a high degree of ligand-induced protein conformational change and hence flexibility (Rauh et al., 2004; Andersen et al., 2000). Several studies have tried to catalogue the degree of burial of active sites and have offered some generalizations relating catalytic activity with degree of ligand burial (Ben-Shimon et al., 2005). Other works focus on detecting crevices and grooves on protein surfaces, as an indirect way of locating ligand binding sites or active sites (Mitchell et al., 2001; Kawabata et al., 2007). In protein-protein complexes, on the other hand, the constituent monomers oftentimes overlap physically in the bound form, resulting from induced fit mechanism facilitated by conformational changes due to protein flexibility (Goh et al., 2004; Bui et al., 2006). There is a limit to such overlap, however: generally, the hydrophobic cores of the individual monomers are very seldom breached (Kimura et al., 2001). To the best of our knowledge, prior to this study there have been no numerical measures of LBS burial available that do not depend in protein size (i.e., relative), and hence no way to directly and objectively compare LBS depths in different proteins, although various recent works have succeeded to some degree (Tan K.P., et al. 2011; Coleman R.G. & Sharp K.A., 2010; Kalidas Y. & Chandra N., 2008; Coleman R.G. & Sharp K.A., 2006; Varrazzo D. et al., 2005; Tan K.P. et al., 2013).

Here we describe two complementary computational procedures we call the "cutting plane" and "tangent sphere" methods. The two methods are generally applicable to protein-ligand interactions (PLI) when the coordinates of residues in the ligand binding site are known, or to protein-protein interactions (PPI) if the coordinates of the residues at the PPI interface are known. In PLI, the methods are used to quantitatively determine the degree of burial of ligand binding sites in proteins (present work; see also and Reyes, V.M., unpublished [b.]), while in PPI, they may be used to quantitatively determine the degree of physical overlap between the monomer partners in the protein complex (Reyes, V.M. (unpublished [c.]). The 'cutting plane' is then defined as that normal to the line formed by the global and local centroids, and containing the latter (Figure 1A). Another name for 'cutting plane' is 'secant plane' but we shall use the former here. The equations of the cutting plane and tangent spheres are derived from the coordinates of LC and GC. From the equation of the CP, the percentage of protein atoms on the 'external side 'of the CP (side opposite the global centroid, purple area) may be calculated exactly and is directly proportional to the degree of burial of the ligand binding site. Likewise, from the equation of the TS, the percentage of protein atoms inside the TS (green area) may be calculated exactly and is inversely proportional to the degree of burial of the ligand binding site. The CP and TS methods find specific use in a procedure we recently developed regarding the prediction of protein-ligand interactions (Reyes, V.M. and Sheth, V.N., 2011; Reyes, V.M., unpublished [a.]) and protein-protein interaction partners (Reyes, V.M. (unpublished [c.]).

## 2 Datasets and Methods

**2.1 The Test Sets.** We have used two test sets in this work: (1.) an artificial protein made up of a 3D cubic lattice of points (whose edge = 1.5 units) in the overall shape of a sphere of radius 50 units, and (2.) a set of 67 representative protein 3D structures used as a dataset in the work of Laskowski et al., 1996. We decided to apply our methods to an artificial protein (above) so that we could study its behavior in as controlled a way as possible.

**2.1.1 Creating the Artificial Protein.** The artificial protein was a grid of points in 3D space in the shape of a sphere with center at the origin and radius 50.0 units (Angstroms). First a 3D cubic grid of points centered at the origin with edges 100 units long ($-50.0 \leq x \leq 50.0$, $-50.0 \leq y \leq 50.0$, $-50.0 \leq z \leq 50.0$) was generated by running a Fortran program that employs a set of three nested do loops, one each for generating the x-, the y-, and the z-coordinates. The grid of points generated are separated by equal distances of 1.5 Å along the x, y and z directions. Then points lying inside the sphere with equation $x^2 + y^2 + z^2 \leq 50.0^2$ are collected. The LBS entrance is assumed to be at the 'north pole', i.e., at the point (0,0,50) and LBS centroids are assumed to all lie along the z- axis from (0,0,50) to (0,0,-50), and equally spaced at a constant separation of 1.0 Å, yielding a total of 101 LBS's in all (including the one at the origin). Thus, 101 artificial proteins are created, all in the form of a spherical grid of points centered at the origin with radius 50, but differing in the location of their LBS's: the first with LBS at (0,0,50), the shallowest LBS; the second with LBS at (0,0,49), the second shallowest LBS; the third with LBS at (0,0,48), the tgird shallowest LBS; …; the 99th with LBS at (0,0,-48), the third deepest LBS; the 100th with LBS at (0,0,-49), the second deepest LBS; and the 101th with LBS at (0,0,-50), the deepest LBS. These 101 artificial proteins were then analyzed using the CP and TS methods. `

**2.1.2 The Laskowski Data Set.** This data set is composed of 67 single-chain monomeric enzymes with bound ligands, some of which contain single and others multiple bound ligands of various sizes. These proteins together with the ligands they contain are shown in Table 1 of Laskowski, et al., 1996. These proteins were selected from the 1995 PDB release based on their E.C. classifications, such that all E.C. numbers are represented once, the one with the highest resolution and best R-factor being selected. Although all are single-chain, 31 have a single domain, 30 have two domains, five have three and one has four. These 67 test structures contain varying numbers of bound ligands (from none to 11). Each was down into 1:1 protein-ligand complexes. For example, if the protein P in the structure has 4 different ligands W, X, Y and Z bound, it is broken down into four protein-ligand complexes, namely, P-W, P-X, P-Y and P-Z, corresponding to four LBSs. Thus, in all, a total of 184 LBS's were derived from this set of 67 test protein structrues. It is this set of 184 LBS's that we test in this work.

**2.2 The 'Cutting Plane' Method**. The cutting plane method (together with the tangent sphere method) is illustrated in Figure 1A. From the coordinates of the amino acid residues making up the LBS, the LBS centroid, a local centroid (LC), may be calculated. On the other hand, the protein molecular centroid, the global centroid (GC), may be calculated from the PDB coordinate file. From these two points, the equation of the cutting plane in the form Ax + By + Cz + D = 0 may be determined by application of simple vector algebra as follows:

Let the local and the global centroids be the points L(p,q,r) and G(s,t,u). The CP is normal to line segment LG and contains L. Thus vector $\overrightarrow{LG}$ = <s-p, t-q, u-r> is normal to the CP. For any point Q:(x,y,z) on the CP, vector $\overrightarrow{LG}$ must be normal to vector $\overrightarrow{LQ}$ = <x-p, y-q, z-r>, which means their dot product is zero: $|LG||LQ|\cos 90° = 0$. Since $\overrightarrow{LG} \bullet \overrightarrow{LQ}$ = (s-p)(x-p) + (t-q)(y-q) + (u-r)(z-r) = 0, we have, upon rearrangement: (s-p)x + (t-q)y + (u-r)z + D = 0 or Ax + By + Cz + D = 0, where A = s-p, B = t-q, C = u-r and D = p(p-s) + q(q-t) + r(r-u). It can be readily shown that the global centroid is always on the positive side of the CP; hence, the percentage of protein atoms on the exterior side of the CP is readily calculated by setting $Ax + By + Cz + D \leq 0$. In Figure 1B, the behavior of the CP index (CPi; the % of protein atoms on the external side of the CP) is monitored as the LBS goes from very shallow to very deep (purple areas); the CPi increases from 0% to 100%.

**2.3 The Tangent Sphere Method.** The tangent sphere method (together with the cutting plane method) is illustrated in Figure 1A. Knowing the coordinates of the LBS centroid (the local centroid, LC) and the protein molecular centroid (the global centroid), the equation of the tangent sphere can be readily determined since its center is at GC and its radius is the distance between LC and GC. The percentage of protein atoms inside the TS may then be readily calculated by setting $x^2 + y^2 + z^2 \leq r$, where r = distance between GC and LC. As with the CP method, the TS method was done in three different ways, namely the 'sidechain submethod'; the 'residue submethod'; and the 'ligand submethod.' And similarly, results from these three submethods were very similar to each other, implying again that the submethod used does not affect the results. In Figure 1B, the behavior of the TS index (TSi; the % of protein atoms inside the TS) is monitored as the LBS goes from very shallow to very deep (green areas); the TSi decreases from 100% to 0%.

**2.4 The CM-CM Method.** This method is just an auxiliary method we tried to implement as an independent way to cross-check the TS and CP methods. This method is simply the determination of the distance between the GC and the LC, in Angstroms, and is a value that is expected to be inversely proportional to the LBS depth, just like the TSi. However, unlike CPi and TSi, this method does not take into consideration the percentage of protein atoms relative to the entire protein; hence this method is dependent upon the size of the protein, and is not reliable to use when the proteins being compared vary dramatically in sizes. On the contrary, CPi and TSi are directly comparable across proteins of widely varying sizes, since they are %'s of protein atoms relative to the whole.

For thoroughness, we performed CPM and TSM in three slightly different ways depending on how the LC coordinates are calculated, namely: (1.) based on the coordinates of the atoms of the amino acid residues that contact the ligand; we designate this the 'sidechain submethod'; (2.) based on the coordinates of *all* the atoms of the amino acid residues that contact the ligand; we designate this method as the residue submethod': and (3.) based on the coordinates of the ligand itself, which we designate as the 'ligand submethod'. In all cases, results from these three submethods were very similar to each other, so essentially it may be concluded that the submethod used does not affect the results (data not shown).

## 3 Results and Discussion

**3.1 Theoretical Case: The Artificial Protein.** The behavior of the artificial protein under the CP and TS methods is illustrated in Figures 2, panels A, B and C. Binding sites lie along the z-axis in 1 unit increments, and the opening of the LBS is at the top, at point (0,0,50) (Figure 2A). Starting from the upper left-hand corner and following the small red arrows, we go from the artificial protein with the shallowest LBS at (0,0,50), to the one with the deepest LBS at (0,0,-50). Midway between these two extremes is the case where the LBS coincides with the protein molecular centroid (at (0,0,0). In the first half of the graph, from the first to the 50$^{th}$ LBS, we note that TSi starts from 100% and decreases quickly then more slowly until it reaches a minimum at 0% (Figure 2B, blue data points); meanwhile, CPi starts from 0% and increases to a maximum at 50% (Figure 2B, pink data points). The other half of the graph, that from 51st to 101th LBS, is a mirror image of the first half, from x=1 to x=50. In this half, TSi increases from 0% to 100%, first slowly and then quickly (Figure 2B, blue data points), while CPi decreases from 50% to 0% (Figure 2B, pink data points). Note that in this case, the location of the LBS opening is known to be at (0,0,50). In this case, we can define the "anterior half" and the "posterior half" of the protein. Imagine the CP passing through the molecular centroid and cutting the protein in half; the half containing the LBS opening is the "anterior half" of the protein, whole the other half (not containing the LBS opening) is its "posterior half."

Had the location of the LBS opening not been known, or known but not taken into consideration, there would have been an ambiguity in the locations of the individual LBS's as to whether they are in the anterior or posterior half of the protein. In that case, the proteins with the shallowest to deepest LBS's would have been determined based solely on their TS indices (largest to smallest) and CP indices (smallest to largest). Thus the ambiguity between the anterior and posterior halves would have caused the right half of the plot in Figure 2B (corresponding to the posterior have) to merge with the left half (corresponding to the anterior half) and the plot would have been that on Figure 2C. The TSM and CPM plots for the 67 test proteins (to be discussed shortly) in Figures 4A and 4B, respectively, correspond to this situation.

**3.2 Application to the Laskowski Protein Dataset.** In applying the CP, TS and CM-to-CM methods we set arbitrary limits for "shallow", "medium" (intermediate depth) and "deep" LBS predictions. For the CP method, they are: (%'s refer to % of protein atoms on the external side of CP): 0.0%-10.0% for shallow; 10.1%-20.0% for medium; and 20.1%-42.0% for deep LBS's (42% is a close upper bound for the results). For the TS method, they are ((%'s refer to % of protein atoms inside the TS): 100.0%-40.1% for shallow; 40.0%-10.1% for medium; and 10.0%-0.0% for deep LBS's. For the CM-to-CM method, they are (Å's refer to the distance between the global centroid, GC, and the local centroid, LC): 38.0Å - 15.1Å for shallow; 15.0 Å-10.1Å for medium; and 10.0Å -0.0Å for deep LBS's. These were the best limit values as judged from the frequency distributions obtained, as discussed below.

**3.2.1 The Tangent Sphere Method as Applied to the Laskowski Dataset.** We first determined the frequency distribution of the prediction results (i.e., "shallow", "intermediate" and "deep") from the TS method, using the ligand, residue and sidechain submethods described earlier. From the distributions (Figure 3A.1, 3A.2 and 3A.3, respectively), it may be concluded that the TS method has more resolving power at the lower bins (shallow LBS's) than at the higher bins (deep LBS's), indicating that it has higher specificity for shallow LBS's than to deep LBS's.

Figure 4A shows the results from the TS method from the 3 submethods: ligand (blue data points), residue (pink data points) and sidechain (yellow data points) submethods. All show the same trend in going from shallow to deep LBS's: decreasing rather quickly in the first half, then more slowly on the second half, almost plateauing near 0% towards the deepest LBS's. We note that this is very similar to the behavior of the artificial protein in Figure 2C (blue data points) and the first half of the plot in Figure 2B (blue data points).

Let us go back to the concept of "anterior" and "posterior" halves of a protein. These are both illustrated in the central object in Figure 1B. In the case of the artificial protein, the researcher knows the exact location of the LBS in relative to the LBS opening, hence it is possible to accurately arrange the proteins from those that have shallowest LBS's (in the anterior half) to those that have deepest LBS's (in the posterior half). In actual work as well in our application to the Laskowski dataset, however, the LBS opening is either unknown or known but not taken in consideration due to automation of the process. Thus in these cases, it is not possible to differentiate the LBS's lying in the anterior half from those that lie in the posterior half. Thus there is such ambiguity in the case of the Laskowski test proteins (where the LBS opening may have been known, but not taken it into account because of automation).

**3.2.2 The Cutting Plane Method as Applied to the Laskowski Dataset.** As in the TS method, we determined the frequency distribution of the prediction results (i.e., "shallow", "intermediate" and "deep") using the ligand, residue and sidechain submethods described earlier. From the distributions (Figure 3B.1, 3B.2 and 3B.3, respectively), it may concluded that the CP method has more resolving power at the higher bins (deep LBS's) than at the lower bins (shallow LBS's), indicating that it has higher specificity for deep LBS's than to shallow LBS's. This is opposite behavior to that of the TS method.

Figure 4B shows the results from the CP method from the 3 submethods: ligand (blue data points), residue (pink data points) and sidechain (yellow data points) submethods. All show the same trend in going from shallow to deep LBS's: increasing rather slowly at first, then faster and almost linearly thereafter, towards the deeper LBS's, and then stops at between 39% and 44% at the deepest LBS's. We note that this is very similar to the behavior of the artificial protein in Figure 2C (pink data points) and the first half of the plot in Figure 2B (pink data points).

As was explained in the previous section, there is an ambiguity in the latter case. However, as can be seen from Figure 4A, the CP results do not reach 50 but stops between 39%, 40% and 44% for the sidechain, ligand and residue submethods, respectively. This means that the LBS's do not reach the depth of the protein molecular centroid. Therefore it is reasonable to conclude that there are no LBS's on the posterior half of the protein, i.e., all LBS's are in the anterior half. This must be so since it is quite improbable that no LBS's are found on and around the global centroid while LBS's on the posterior half exist. As for the case of the TS method in Figure 4B, we see that it does reach 0% near the deepest LBS's (0.0%, 0.15% and 0.25% for the ligand, residue, and sidechain methods, respectively). However, as was discussed earlier, the TS method has low resolving power for the deep LBS's and in fact the curve plateaus near at 0% near the deepest LBS's. The TS method seems unable to differentiate between LBS depths when the TS reaches a radius of within 5 Å from the global centroid. This appears to validate our conclusion that there are probably no LBS's on the posterior halves of the 67 test proteins.

**3.2.3 The CM-to-CM Method.** Figure 3C.1, 3C.2 and 3C.3 show the frequency distribution of the results from the CM-to-CM method, using the ligand submethod, residue submethod and sidechain submethod, respectively. Here, the horizontal axis is the distance in Angstroms between the local CM and the global CM ("CM-to-CM value"), going from shallowest LBS at 38 Å on the extreme left, to deepest LBS at 0 Å on the extreme right, partitioned into intervals of width 1.9 Å, for a total number of intervals ("bins") of 20. The vertical axis is the number of LBS's (out of the 184 total) with CM-to-CM values corresponding to each interval or bin along the x-axis. For example, in the CM-to-CM method, ligand submethod, there is 1 LBS with a CM-to-CM value between 38 Å and 36.1 Å (first bin), while there are no LBS's with a CM-to-CM value between 1.9 Å and 0% (last bin). Note that the resolving power of the CM-to-CM method is highest on the first half of the plot (shallow LBS's), dipping to a minimum about halfway through the second half, and then increases again towards the end of the second half (very deep LBS's), indicating that it has higher specificity for shallow to almost-shallow LBS's and for very deep LBS's.

Figure 4C shows the results from the CM-to-CM method from the 3 submethods: ligand (blue data points), residue (pink data points) and sidechain (yellow data points) methods. All show the same trend in going from shallow to deep LBS's: decreasing rather quickly in the first quarter, then more slowly on the second and third quarters, then more slowly again and almost plateauing near 0% towards the deepest LBS's. That there are most probably no LBS's on the posterior halves of the 67 test proteins is also supported by these CM-to-CM data since the LC to GC distance does not reach 0 Å (3.0 Å, 1.9 Å and 2.4 Å for the ligand, residue and sidechain submethods, respectively.) . This implies, as in the CP method, that no LBS is close to the GC, hence there are probably no LBS's in the posteriori half of the proteins.

**3.3 Summary of Final Results.** From the 67 Laskowski dataset of protein structures, there are a total of 184 LBS's since some of the proteins in the dataset have more than one ligand (from as low as none to as many as 11; see Table 1 and legend), as mentioned earlier. The determination of the mutual intersections of the shallow, medium (intermediate depth) and deep predictions by the TSM, CPM and CM-CM method, using the 3 submethods (ligand, residue and sidechain submethods), respectively, have been determined, and the results are shown and summarized in the Venn Diagrams in Figure 5, panels A, B and C for the TSM, CPM and CM-CM methods, respectively, and similarly in Figure 6, panels A, B and C. The overall results from the TSM, CPM and CM-CM are presented in combined form in Figure 7, Panel A while in Panel B, only the combined results from the TSM and CPM are presented (with the CM-CM results excluded).

Specifically, Figure 5A, B and C show the first step in the analysis process, which is the pairwise determination of the intersections among the three submethods (ligand, residue and sidechain) in the TSM, CPM and CM-CM methods, respectively. Meanwhile, Figure 6A, B and C show the second step in the process, which is the determination of the mutual 3-way intersections among the three submethods (ligand, residue and sidechain) in the TSM, CPM and CM-CM methods, respectively. Figure 7 shows the final step in the analysis process. In Panel A, the determination of the mutual three-way intersections among the TSM, CPM and CM-CM methods are shown. In this case, 37 structures are common to the TS, CP and CM-to-CM shallow LBS predictions, 8 are common to the medium LBS predictions, and finally 47 are common to the deep LBS predictions. The identities of these 37 shallow, 8 medium and 47 deep LBS's are shown in Table 2. Thus a total of 37+8+47 = 92 LBS's out of the 184 LBS's, or 50%, were predicted identically by our three methods, TS, CP and CM-CM. Panel B is the same as Panel A except that the CM-CM results are excluded, as this method is disfavored relative to the TSM and CPM methods because it (the CM-CM method) is dependent on protein size, while the TSM and CPM are not. In this case, 38 structures are common to the TSM and CPM shallow LBS predictions, 11 are common to the medium LBS predictions, and finally 51 are common to the deep LBS predictions. The identities of these 38 shallow, 11 medium and 51 deep LBS's may be gleaned from Table 2. Thus a total of 38+11+51 = 100 LBS's out of the 184 LBS's, or 54%, were predicted identically by the two methods about which this work is about, the TS and CP methods. Note that an additional 1, 3 and 4 LBSs are included in the shallow, medium and deep predictions, respectively compared to the previous tabulation. These additional LBSs represent the ones which were supposed to have been incorrectly predicted by the CM-CM method due to its dependence on protein size.

Finally in Table 3, panels A, B and C, we show the identities and the corresponding CPi and TSi values of the final predictions in Figure 7A and B. Panel A show the deep predictions, panel B the medium predictions, and panel C the shallow predictions. Note that the CPi's of the LBSs in panels A, B and C decrease in that order, while the TSi's increase in the same order. These results demonstrate the complementarity between these two LBS burial depth metrics. To the best of our knowledge, prior to this study, no such (numerical) metrics existed for LBS burial depth. Wirth these two new metrics (TSi and CPi), LBS burial depth in different proteins, regardless of size variations among them, may be compared.

The results we obtained above are very similar to the results described by Laskowski et al.. (1996), wherein the ligand depths were determined by manual (visual) inspection. Hence we conclude that the Cutting Plane (CP) and Tangent Sphere (TS) methods are quite effective and reliable methods of determining and quantifying ligand or LBS burial in proteins, provided that the coordinates of the LBS (local site) and that of the entire protein molecule (global structure) are available.

**3.4 Advantages and Limitations of the Methods.** Prior to this study, there were no numerical measures of LBS burial available that do not depend in protein size, and hence no way to directly and objectively compare LBS depths in different proteins. The CP and TS methods are two complementary methods that provide such a way. The fact that the TS and CP indices are complementary does not mean that they are redundant in the sense that if one is known, the other may be precisely calculated; that is only true for a perfectly spherical protein, but real proteins are never perfectly spherical. That being said, the CP and TS methods would work best if the test protein is roughly spherical, or globular, in shape. If the protein has an overall concavity, or is otherwise irregular in shape, the method becomes less accurate. Thus, we can expect the CP and TS methods to be less accurate for rod-shaped proteins, etc. But since the majority of structurally solved proteins (about 80% of them) are globular ("spheroproteins"), it may be concluded that the CP and TS methods are generally applicable. The complementarity of the two methods (i.e., the CP index is directly proportional to LBS burial depth, while the TS index is inversely proportional to it) is an advantage of employing both methods in parallel since they cross-validate each other. Although TSM and CPM were created out of a need for validating our own results from our prediction studies (Reyes, V.M., unpublished [a.], [b.] and [c.]), it is generally applicable to **any** protein whose atomic coordinates, as well as those of the relevant local sites in it (i.e., LBSs), are known.

## 4 Conclusions.

We have developed a couple of complementary methods, as well as an auxiliary method, to quantify the depth of burial of a ligand or its binding site in a protein structure. These methods, called the "Cutting Plane," the "Tangent Sphere" and the "Centroid-to-Centroid" methods, respectively, are all implemented as Fortran 77/90 programs and are fast, efficient and amenable to high-throughput or big-batch applications. We tested the methods on a set of 67 previously characterized structures containing a combined total of 184 LBS's, as well as on an artificial protein consisting of a grid of points in the shape of a sphere, where LBS's can be arbitrarily assigned anywhere on it. Predictions on the LBSs by the three methods in the form of "shallow", "medium' (intermediate depth) and "deep" are in line with previous characterizations done by human visual inspection. The predictions are also independent of protein size, as they are relative to the size of the entire protein (based on percentages of portions of protein relative to the whole). We believe that these methods may find useful applications in structural proteomics, especially protein dynamics, as ligand burial depth is indicative of the amount of conformational changes the protein undergoes in binding the ligand. To the best of our knowledge, this work is the first attempt at the relative quantification of the depth of burial of ligand or ligand binding sites in proteins using cutting planes and tangent spheres as described in this work.

**Acknowledgments.** This work was supported by an Institutional Research and Academic Career Development Award to the author, NIGMS/NIH grant number GM 68524. The author thanks Ms. Srujana Cheguri, one of his M.S. Bioinformatics graduate students at Rochester Institute of Technology (RIT), for verifying results. The author wishes to acknowledge the San Diego Supercomputer Center, the UCSD Academic Computing Services, and the UCSD Biomedical Library, for the help and support of their staff and personnel. He also acknowledges the Division of Research Computing at RIT, and computing resources from the Dept. of Biological Sciences, College of Science, at RIT.

## FIGURE LEGENDS:

**Figure 1, Panel A: Definition and Illustration of the Cutting Plane and Tangent Sphere Methods**
Illustration of the Cutting Plane and Tangent Sphere Methods and how they complement each other. The brown circle represents a spherical protein, with a ligand binding site whose opening is at the top; the ligand is shown as a small red circle. The blue horizontal line is the cutting plane (CP), cutting the protein through its ligand binding site (LBS) centroid. This centroid is different from the protein molecular centroid, shown at the center of the protein as a ©: the former is a local centroid, while the latter is a global centroid. The tangent sphere (TS) is shown as a green circle tangent to the CP at the LBS centroid. The exterior side of the CP is defined to be the side opposite the protein molecular centroid, white its interior side is that containing the protein molecular centroid. The volume of the protein lying on the CP exterior side is shown in purple, while the volume of the protein lying inside the TS, on the interior side of the CP, is shown in green.
**Panel B: Illustration of the CP and TS Methods on a Artificial Protein in the form of a Sphere.** To start with**,** the protein is shown with a shallow ligand binding site. The CP (blue line) and TS (green circle) are shown as well, and the volume of the protein lying on the exterior of the CP and in the interior of the TS are shown in purple and green, respectively. These volumes, expressed as percentages of the total protein volume, are also shown above (%CP) and below (%TS) the cutting plane, respectively; we designate them as the CP and TS indices, respectively. In the succeeding panels, the LBS is made deeper and deeper into the protein; the behavior of %CP and %TS follow two different trends and are discussed in the text and plotted in Figure 3. In both panels, the V-shaped concavity in the circle represents the LBS, but shown only for purposes of illustration. In all our calculations, the receptor protein is conceptualized as a perfect sphere without any such concavity embodying the LBS.

**Figure 2: Plots of CPM and TSM Results for the Theoretical Case ("Artificial Protein")**
**Panel A:** The artificial protein is shown as a spherical grid of points with center at the origin and the LBS assumed to be at the "top" i.e., at its maximum point along the positive z-axis. Plots of the CP and TS indices for this case where the LBS opening is precisely known is shown in Panel B. Plots of the CP and TS indices for the case where the LBS opening is **not** known is shown in Panel C.
**Panel B:** %CP and %TS data from the artificial protein as LBS is made to vary from very shallow (near the surface), to shallow, to deep, to very deep (near the protein centroid), to deepest (near the other end of the protein opposite the LBS opening). The %CP (pink data points) and the %TS (dark blue data points) were plotted against the LBS depth, with the shallowest on the far left and the deepest on the far right. LBS depth exactly midway between the shallowest and deepest depths corresponds to the situation where the LBS is at or very near the protein molecular centroid. See text for a discussion of the %CP and %TS trends as LBS depth varies.
**Panel C:** This is what the plot in Panel A would look like if the LBS opening is unknown in the case of the artificial proteins. The right half of the plot will fold into the left half due to the resulting ambiguity between the anterior and posterior halves of the proteins (see text). This figure shows a basic difference between the artificial proteins and the 67 test proteins. In the artificial proteins the location of the LBS opening is known, and there are LBS's on both the anterior and posterior half of the protein. In the test (real) proteins, the location of the LBS opening is not known, hence there is ambiguity between the anterior half and the posterior half of the proteins in terms of the location of the LBS's. However, data from the CP and CM-to-CM methods indicate that there are no LBS's on the posterior half of the proteins; data from the TS method does not disagree with such conclusion.

**Figure 3: Frequency Distribution Plots for CPM, TSM and CM-CM Methods for the 67 Test Structures**
**Panels A1, A2 and A3:** Frequency distribution of the classification results obtained from the Tangent Sphere method using the three submethods: ligand CM (Panel A1), residue CM (Panel A2), and side chain CM (Panel A3). In the TS method, the LBS depth is measured in terms of the fraction of protein volume lying inside the tangent sphere, and the range of this parameter is (theoretically and experimentally) from 1% to 100% (see text). Each bin or interval on the horizontal axis is 5% in all 3 panels, giving rise to 20 bins. The vertical axes measure the number of structures in each particular bin. Note that TSM has higher resolving power for shallower LBSs than for deeper LBSs.
**Panels B1, B2 and B3:** Frequency distribution of the classification results obtained from the Cutting Plane method using the three submethods: ligand CM (Panel B1), residue CM (Panel B2), and side chain CM (Panel B3). In the CP method, the LBS depth is measured in terms of the fraction of protein volume on the external side of the cutting plane, and the range of this parameter is theoretically from 0% to 50%, but experimentally the upper bound only goes to 42% (see text). Each bin or interval on the horizontal axis is 2.1% in all 3 panels, giving rise to 20 bins. The vertical axes measure the number of structures in each particular bin. Note that CPM has higher resolving power for deeper LBSs than for shallower LBSs.
**Panels C1, C2 and C3:** Frequency distribution of the classification results obtained from the CM-CM method using the three submethods: ligand CM (Panel C1), residue CM (Panel C2), and side chain CM (Panel C3). In the CM-to-CM method, the LBS depth is measured in terms of the distance between the local and the global centroids (LBS and protein CM's, respectively), and the range of this parameter is from 0 Å (local and global CM's coincide), and there is no theoretical upper bound, but in this data set, the upper bound is 38 Å (see text). Each bin or interval on the horizontal axis is 1.9 Å in all 3 panels, giving rise to 20 bins. The vertical axes measure the number of structures in each particular bin. Note that the maximal resolving power of the CM-CM method is intermediate to those of the TS and CP methods.

**Figure 4: Plots of CPM and TSM Results for the 67 Test Structures**
**Panel A:** Results of LBS depth quantification using the CP method done using the 3 submethods, ligand CM (black), residue CM (pink) and side chain CM (yellow). LBS depth is directly proportional to % protein atoms on the external side of the CP, with a lower bound of 0%, and an upper bound of about 42%. Along the horizontal axis are the 184 protein-ligand pairs from the Laskowski dataset arranged from those with lowest CP index (leftmost) to those with highest CP index (rightmost).
**Panel B:** Results of LBS depth quantification using the TS method done using the 3 submethods, ligand CM (black), residue CM (pink) and side chain CM (yellow). LBS depth is inversely proportional to % protein atoms lying inside the TS, with a lower bound of about x%, and an upper bound of about 42%. Along the horizontal axis are the 184 protein-ligand pairs from the Laskowski dataset arranged from those with highest TS index (leftmost) to those with lowest TS index (rightmost).
**Panel C:** Results of LBS depth quantification using the CM-CM method done using the 3 submethods, ligand CM (black), residue CM (pink) and side chain CM (yellow). LBS depth is inversely proportional to the distance between the global and the local centroids, and in this data set, this parameter has a lower bound of about 2 Å and an upper bound of about 38 Å. Along the horizontal axis are the 184 protein-ligand pairs from the Laskowski dataset arranged from those with largest CM-CM distance (leftmost) to those with smallest CM-CM distance (rightmost).

**Figure 5A,B,C: Summary Two-Way Venn Diagrams.** Nine Venn Diagrams showing pairwise intersections of the results from the ligand, residue and side chain submethods for those structures classified as "deep", "medium (intermediate depth), and "shallow." Data in panel A are from the TS method, those in panel B from the CP method, and those in panel C from the CM-CM method.

**Figure 6A,B,C: Summary Three-Way Venn Diagrams.** Three Venn Diagrams showing the mutual (three-way) intersections of the results from the ligand, residue and side chain submethods for those structures classified as "deep", "medium", and "shallow". Data were derived from those in Figure 5, panels A, B and C. The entries in the center of each Venn Diagram are the 3-way intersections among the three submethods (ligand, residue and side chain). Data in panel A are from the TS method, those in panel B from the CP method, and those in panel C from the CM-CM method.

**Figure 7, Panel A: Venn Diagrams Illustrating Commonalities of Prediction by the CP, TS and CM-CM Methods.** The final results from the TS, CP and CM-CM methods of determining LBS burial depth is shown. The top Venn Diagram shows the intersections of the "shallow" predictions by these three methods, and we see that 37 are common among them. The lower left Venn Diagram shows the intersections of the "medium" (i.e., intermediate LBS depth) predictions of these three methods, and we see that 8 are common among them. The lower right Venn Diagram shows the intersections of the "deep" predictions of these three methods, and we see that 47 are common among them. In all 92 cases (=37+8+47), the predictions were identical with those of Laskowski et al. (1996), whose assessments were derived by inspection (visual).
**Panel B:** If the CM-CM results are not considered, we arrive at the Venn Diagram shown here. The common intersection will then be 100 (=38+11+51) instead of 92. The CM-CM method is oftentimes best left out because it is dependent on protein size, since it measures absolute distances in Angstroms instead of percentages of atoms in the protein (relative to that in the entire molecule) as the TS and CP indices do.

**TABLE LEGENDS:**

**Table 1: Tabulation of the 67 Test Structures from the Laskowski Dataset.** Many of the 67 structures (column 1) have more than one bound ligand; each was broken down into 1:1 protein-ligand pairs. For example, consider a structure that has a single protein chain P, with 3 bound ligand molecules, A, B and C. This structure is broken down into P-A, P-B and P-C, and the 1:1 protein-ligand pairs are designated 1, 2 and 3. If a structure contains a single ligand, the protein-ligand pair is designated "0". These are the entries in the third column of the table. Broken down in this way, the 67 structures in the Laskowski data set gave rise to 184 protein-ligand pairs. The second column of the table are the short-hand designations of the 67 structures. For example, "Bl.lig4" would refer to structure with PDB ID "1ELA" and its ligand "ACY" in chain B, residue number 300; "Bl.lbs4" would mean the same, except that the ligand binding site, not the ligand, is the entity on question.

**Table 2: LBS's Judged "Shallow", Medium depth", and "Deep" by the Three Methods.** These are the identities of the 1:1 protein-ligand pairs classified as "shallow", "medium", and "deep" by the 3 methods as shown in the center of each Venn Diagrams in Figure 6, panels A B and C, respectively. The first 3 columns show the results from the TS method, the next three from the CP method, and the last three from the CM-CM method. Each column is the mutual (three-way) intersection of the ligand, residue and side chain submethods for each pertinent classification (deep, medium and shallow) and method (TS, CP and CM-CM).

**Table 3A,B,C: TS and CP Indices of Common Predictions.** The TS indices (first 4 columns), CP indices (next 4 columns) and CM-to-CM distances (last 4 columns) of the Laskowski structures classified as "deep" (panel A), "medium" (panel B), and "shallow" (panel C). In each panel, the three indices from the three submethods (ligand, "lig"; residue, "res"; side chain, "sdc"; average, "ave") are shown in black, and their averages are shown in red. The numbers in blue (8 in all) in the average column of the CM-CM section are the ones that become excluded if the CM-CM method is not considered (see Figure 7A,B), for reasons discussed previously. These indices (the TS and CP indices, in particular.) are precisely the quantitative measures of LBS burial that are referred to in the text. Prior to this study, there were no such numerical measures of LBS burial available, and hence no way to directly and objectively compare LBS depths in different proteins.

# FIGURES:

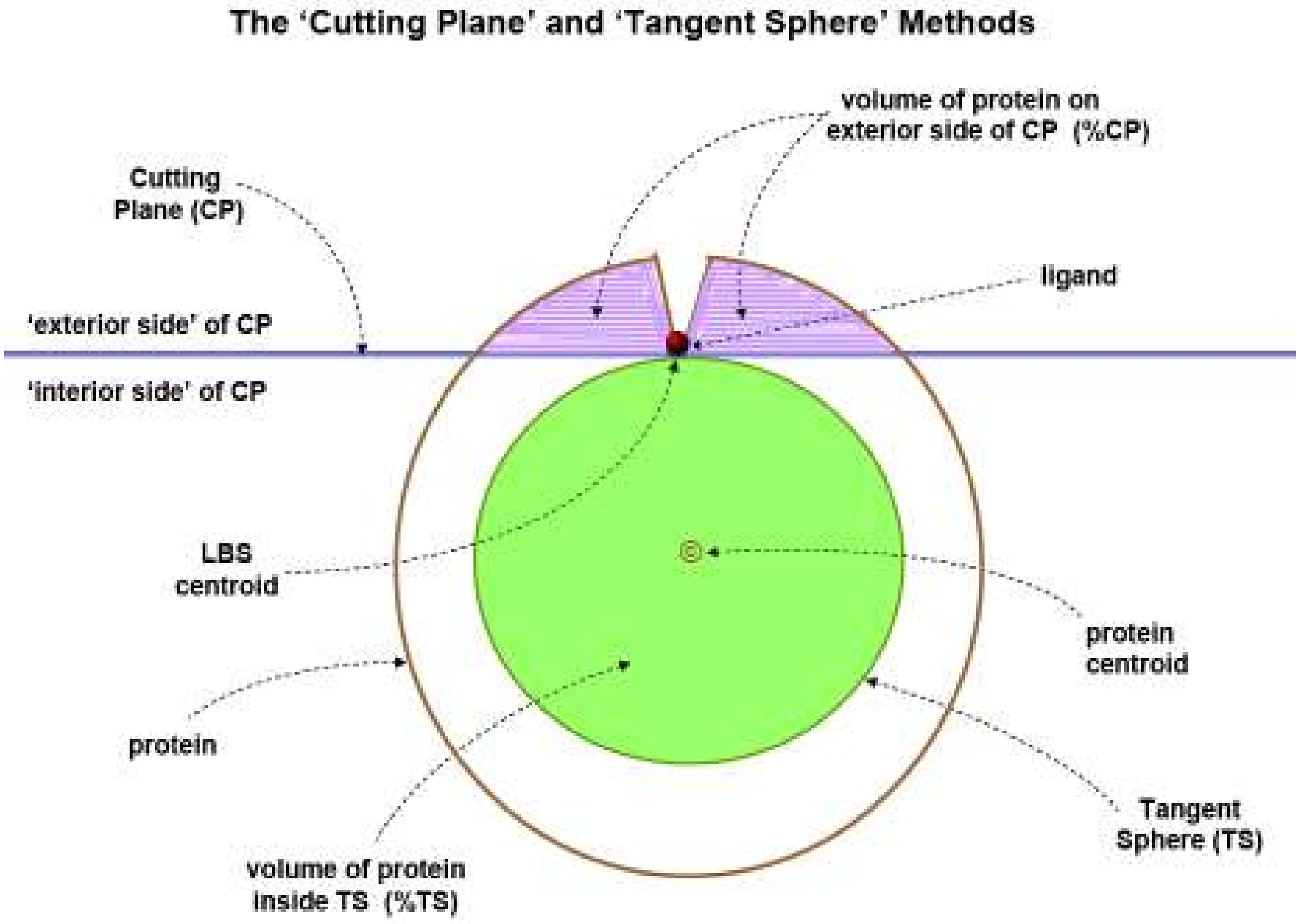

**Figure 1A**

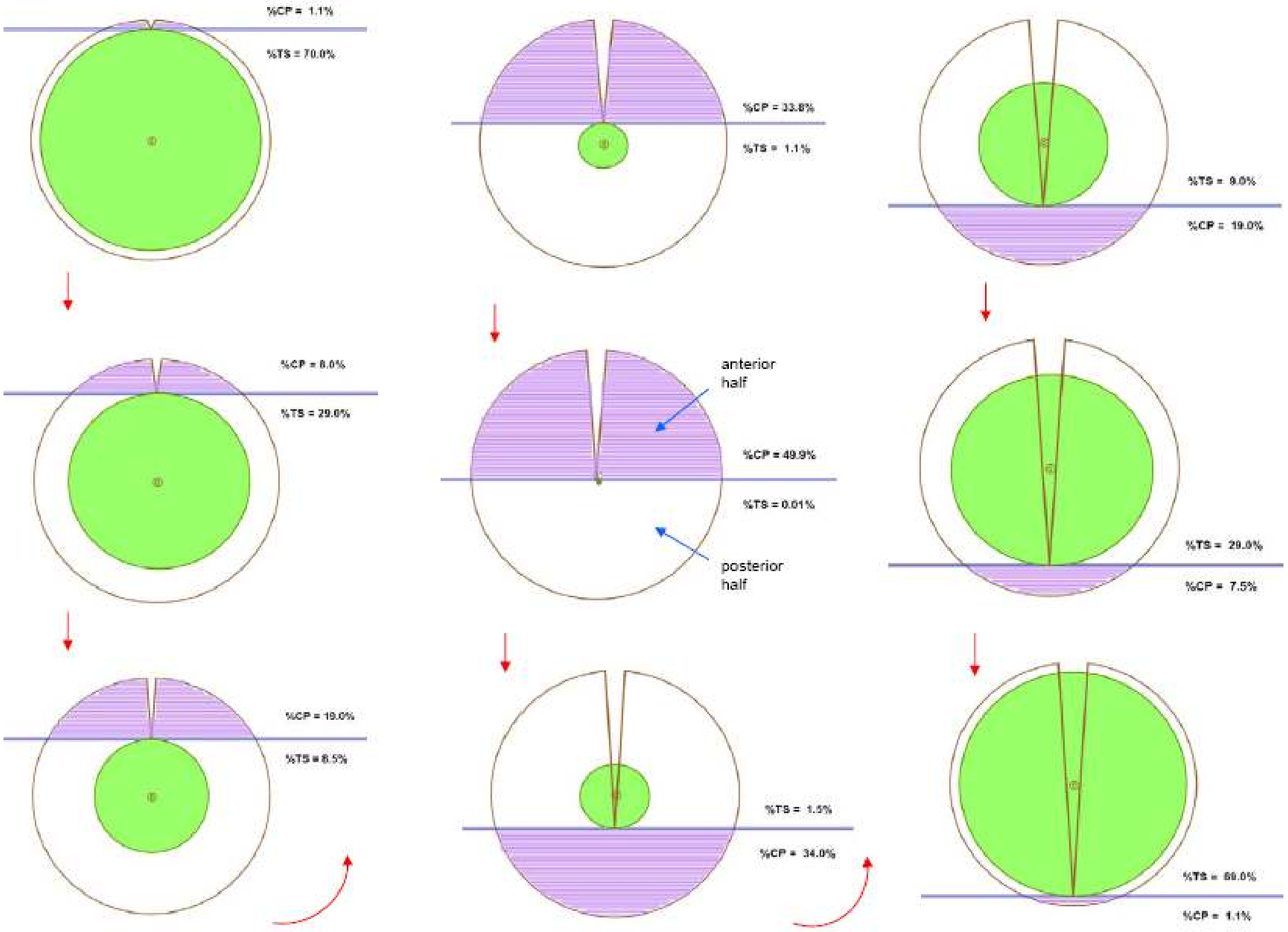

**Figure 1B**

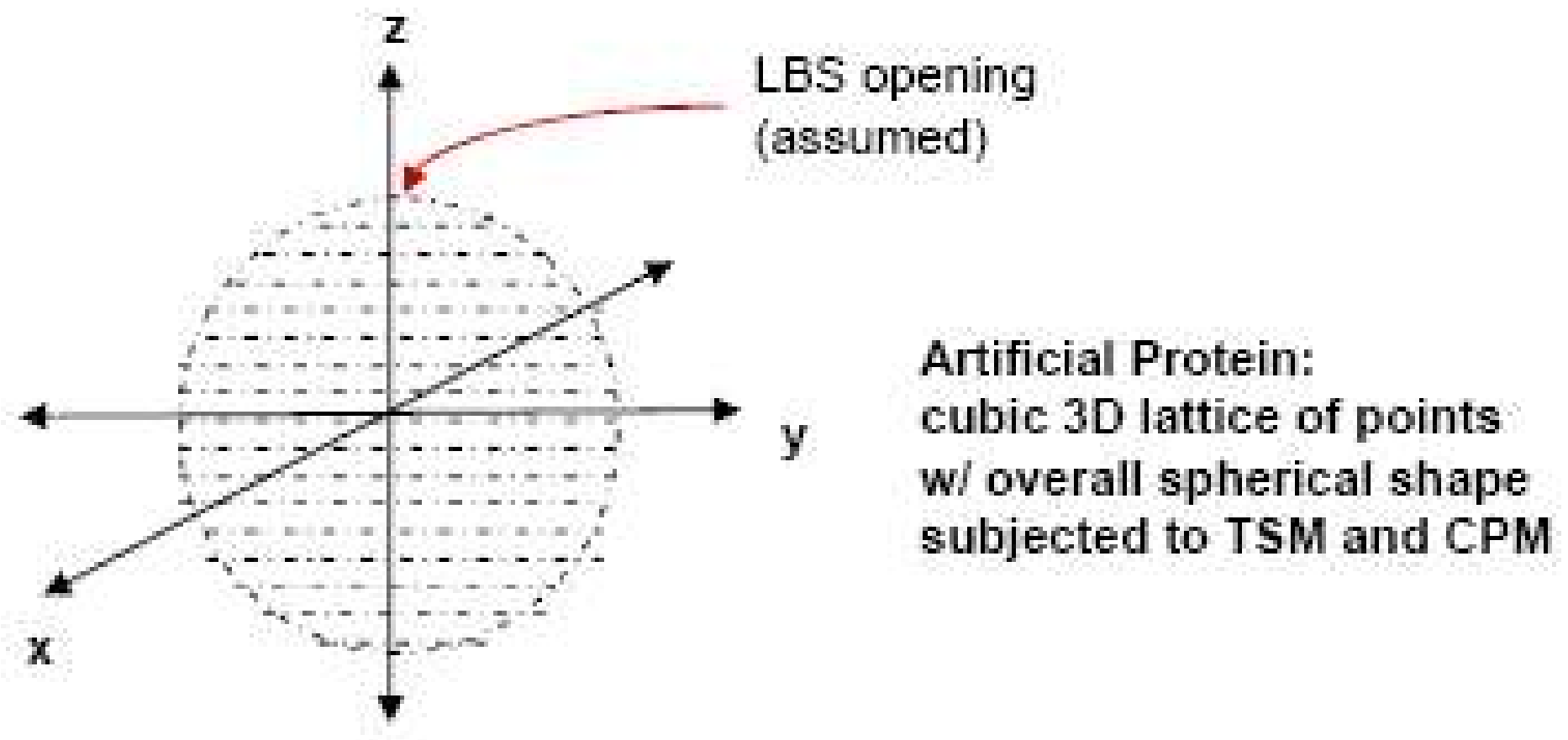

**Figure 2A**

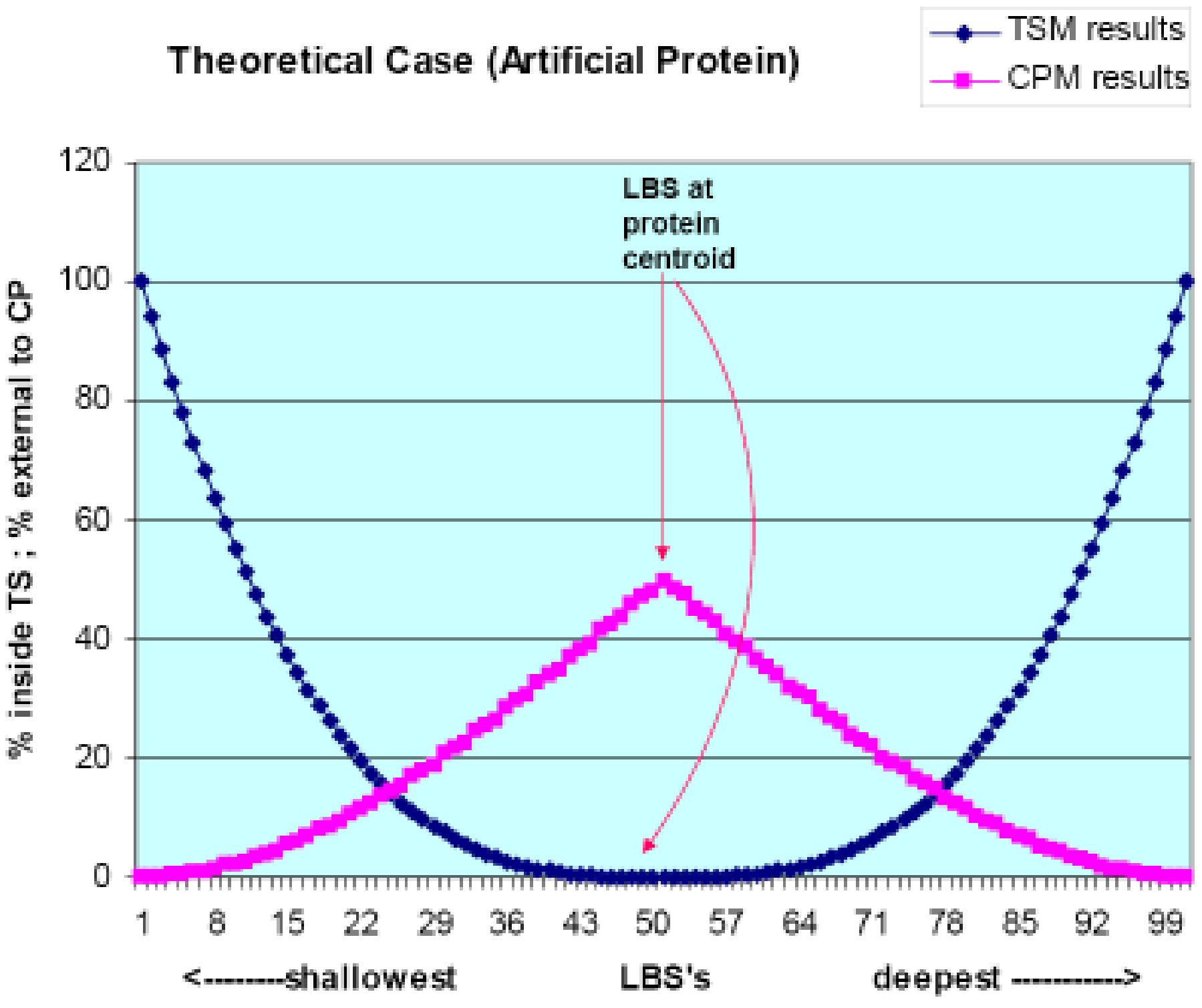

**Figure 2B**

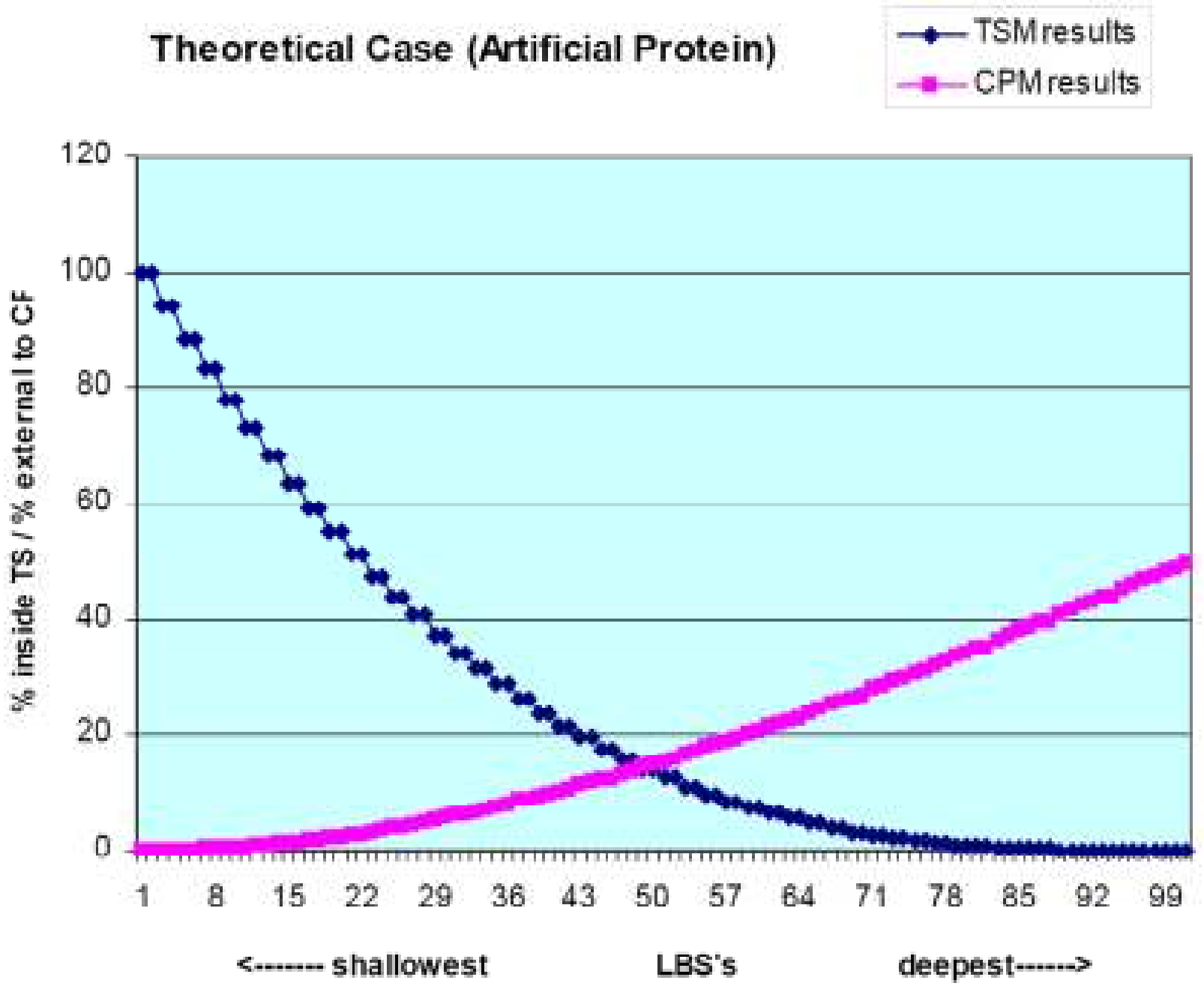

Figure 2C

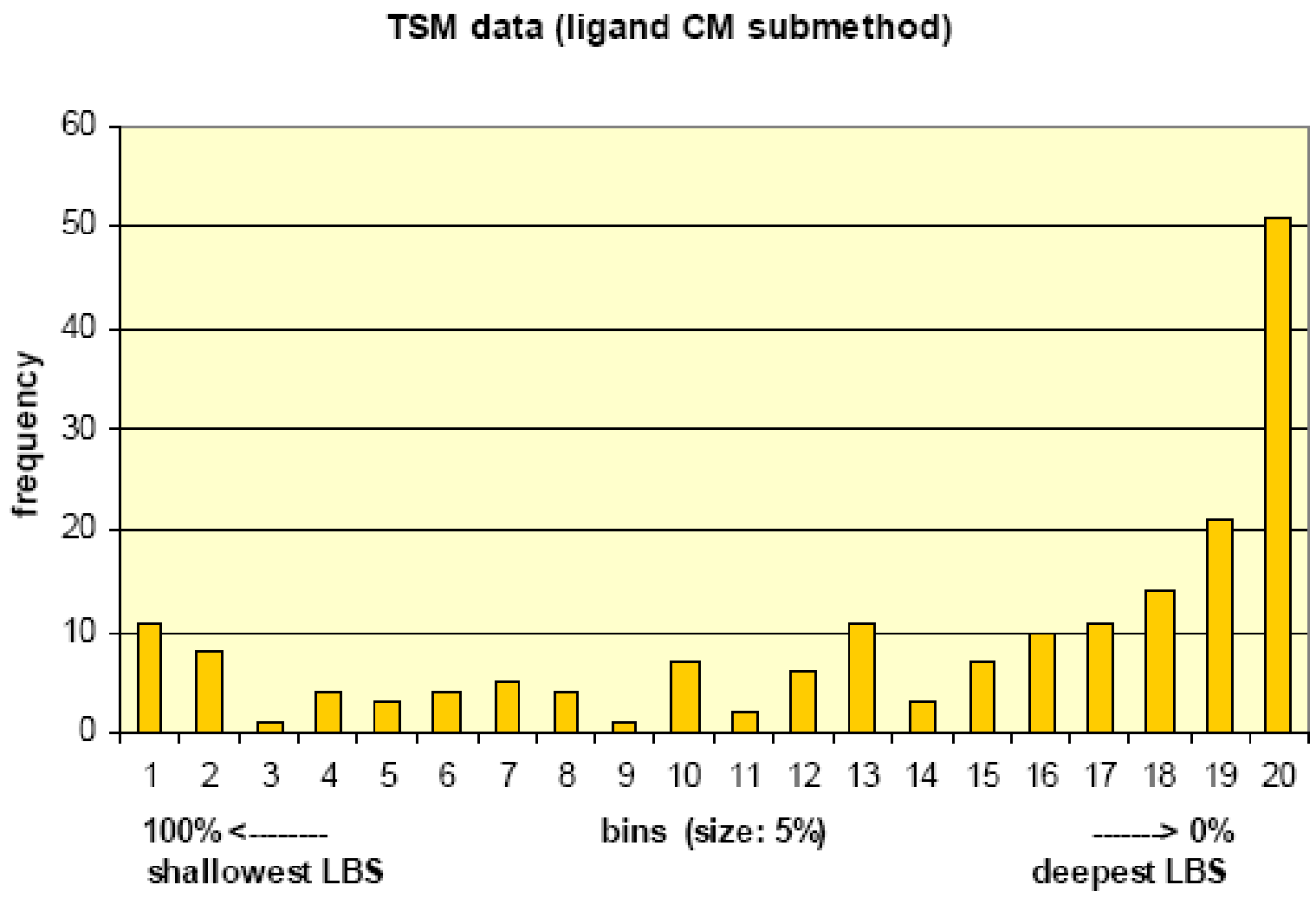

Figure 3A.1

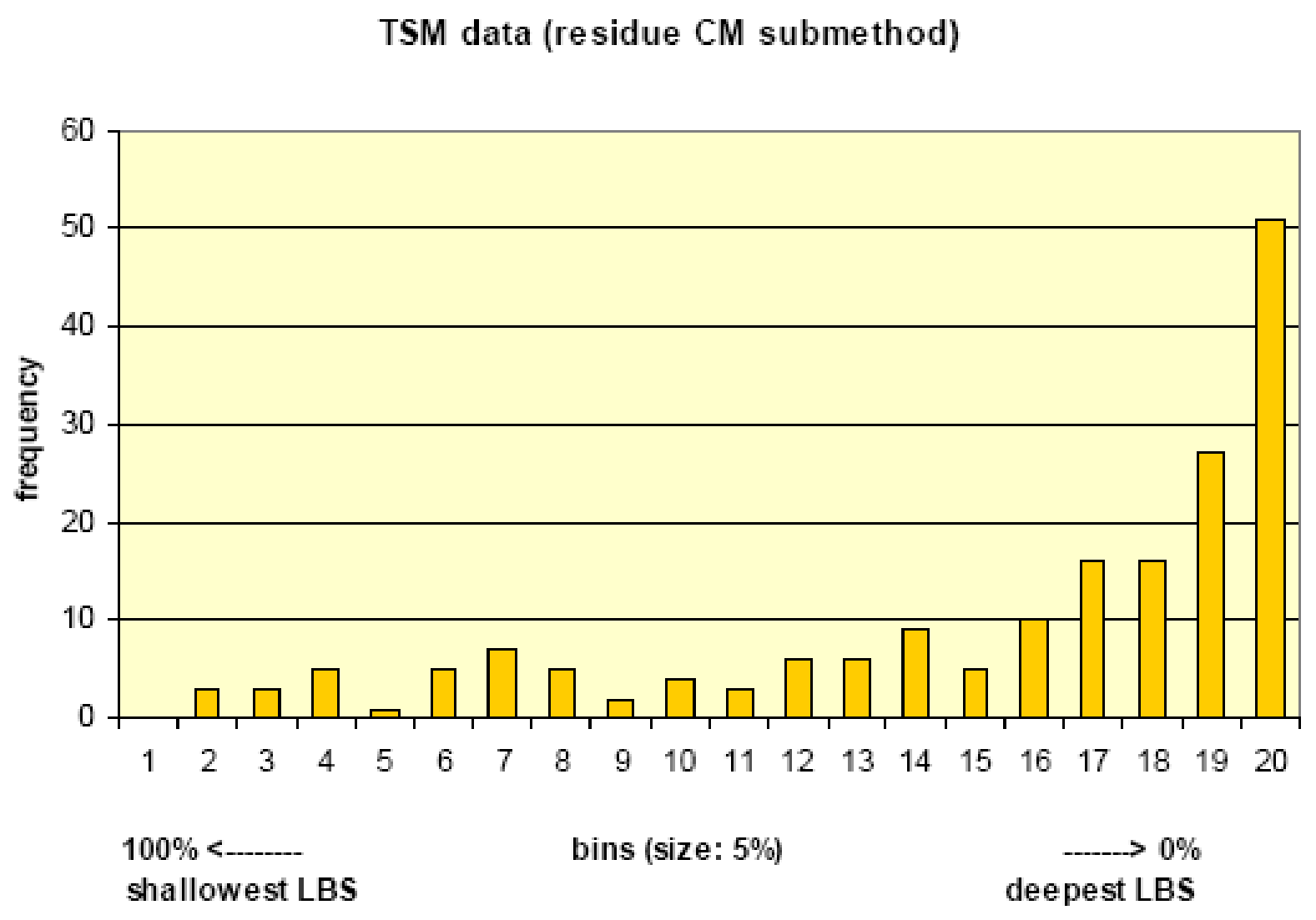

**Figure 3A.2**

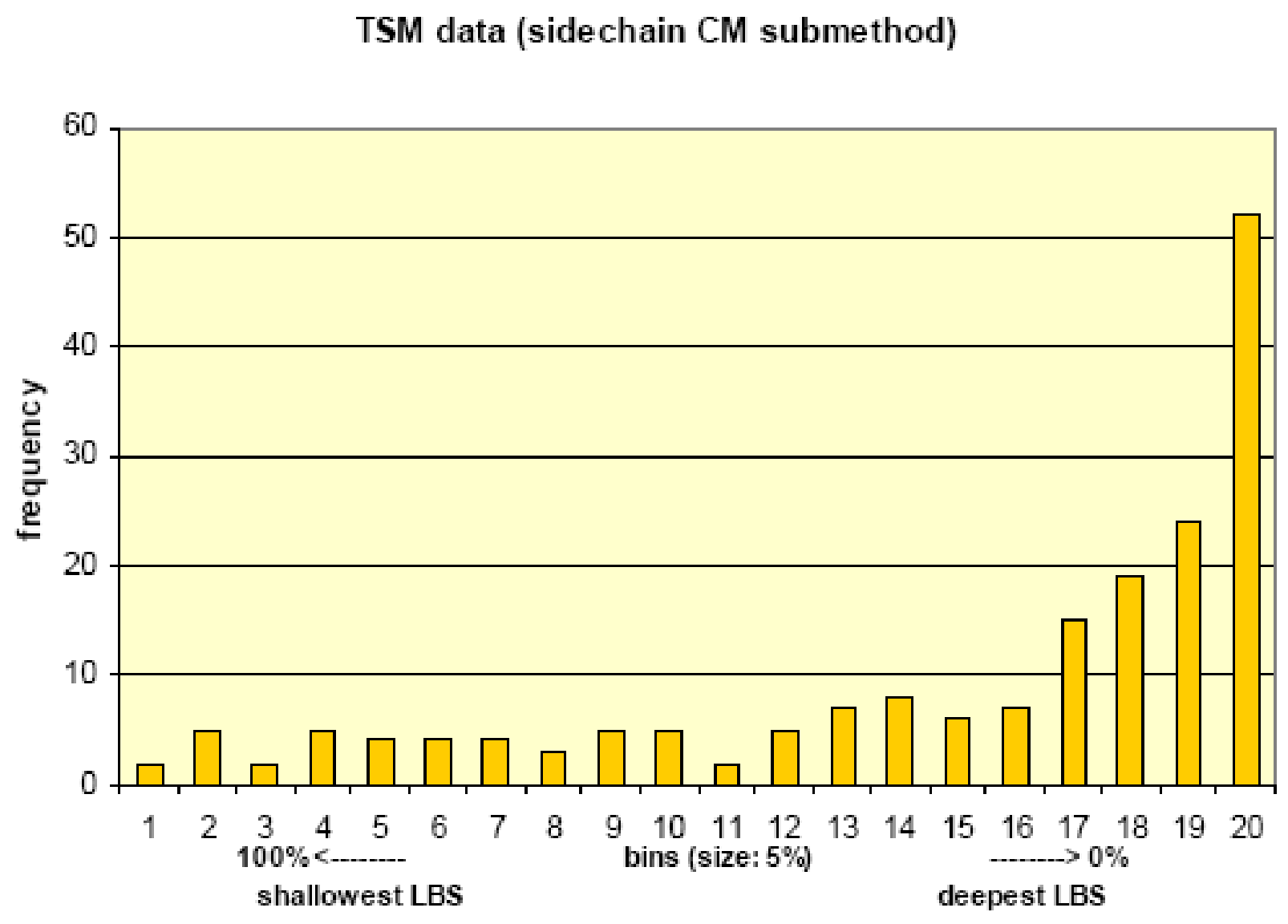

**Figure 3A.3**

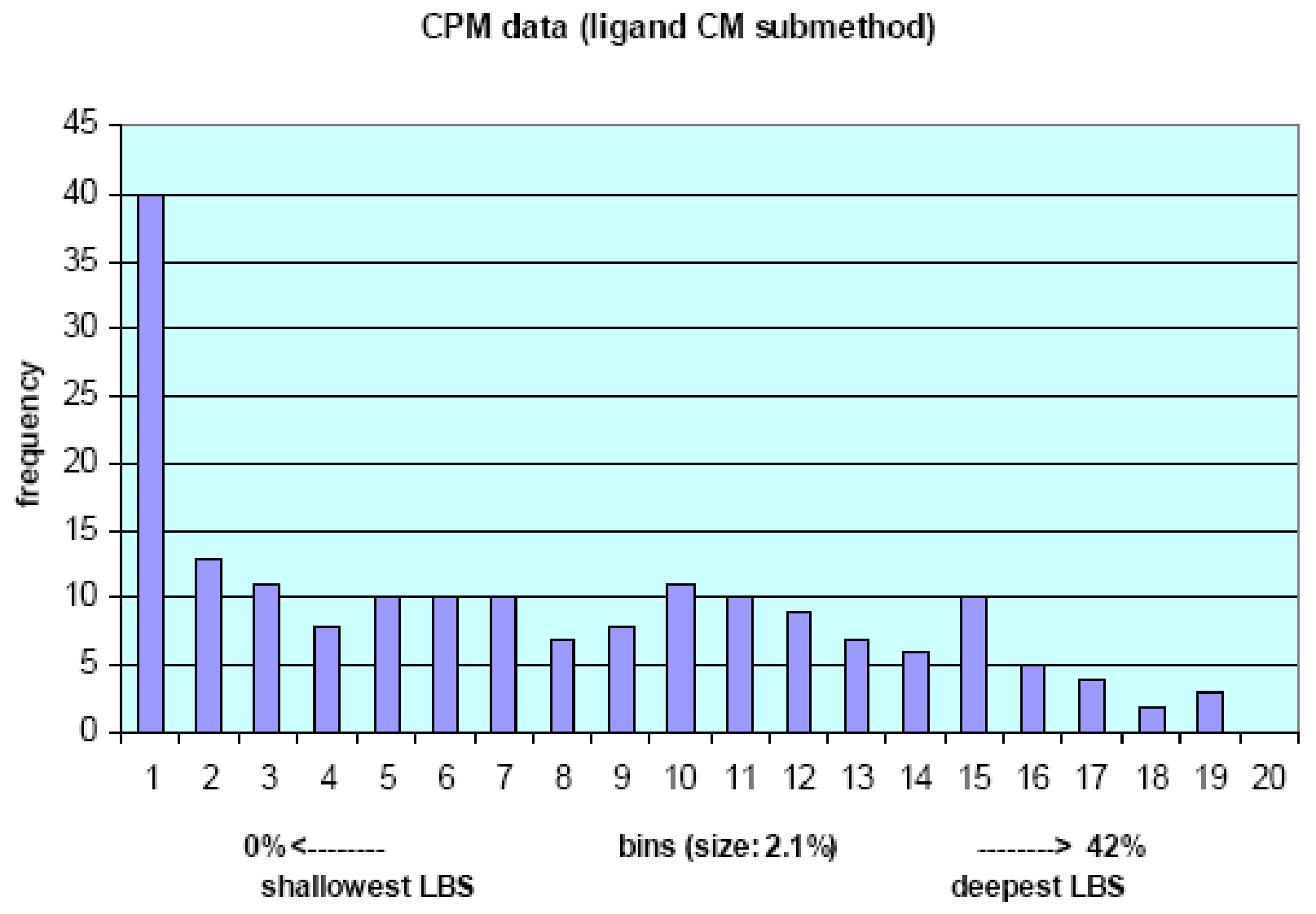

**Figure 3B.1**

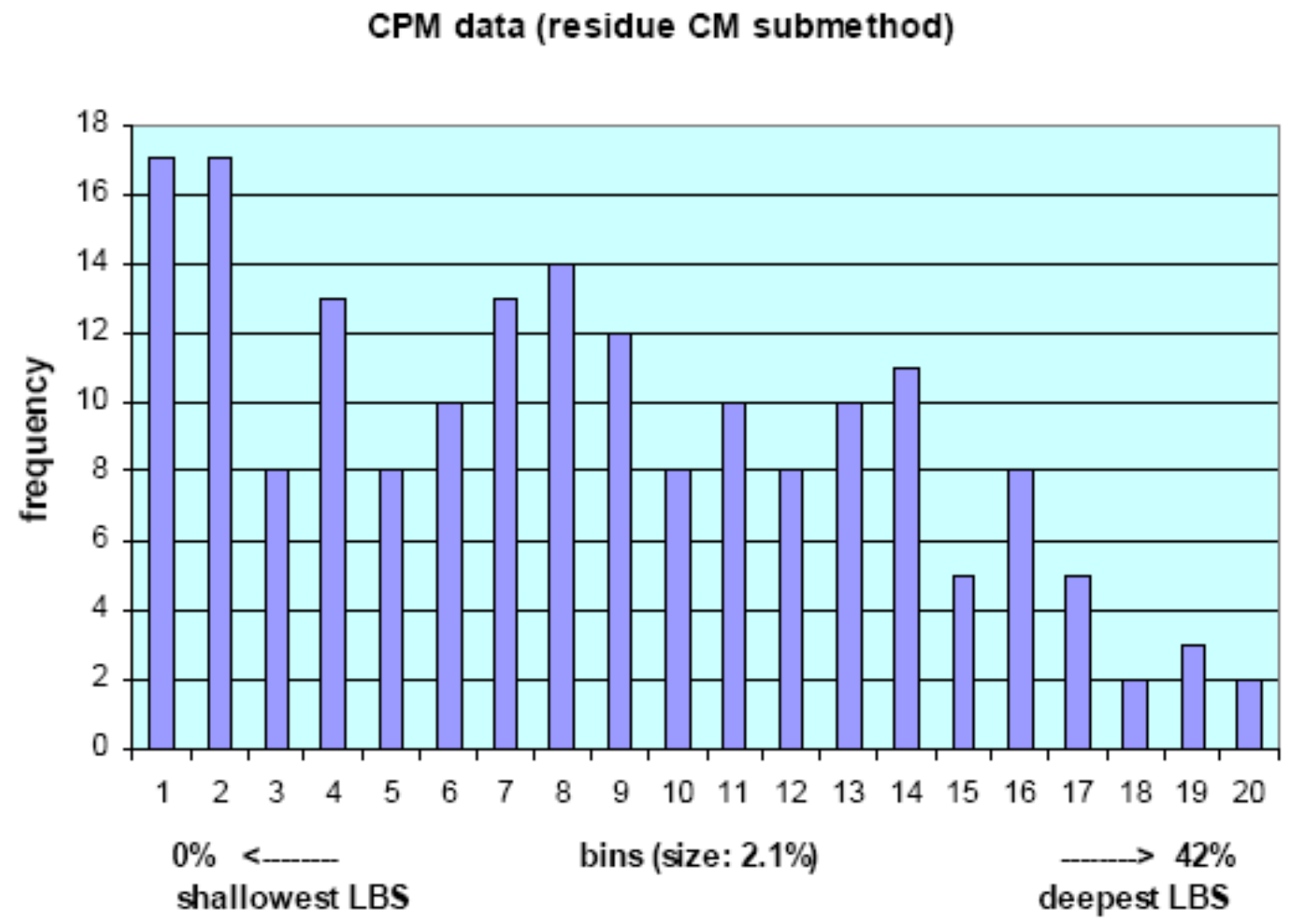

**Figure 3B.2**

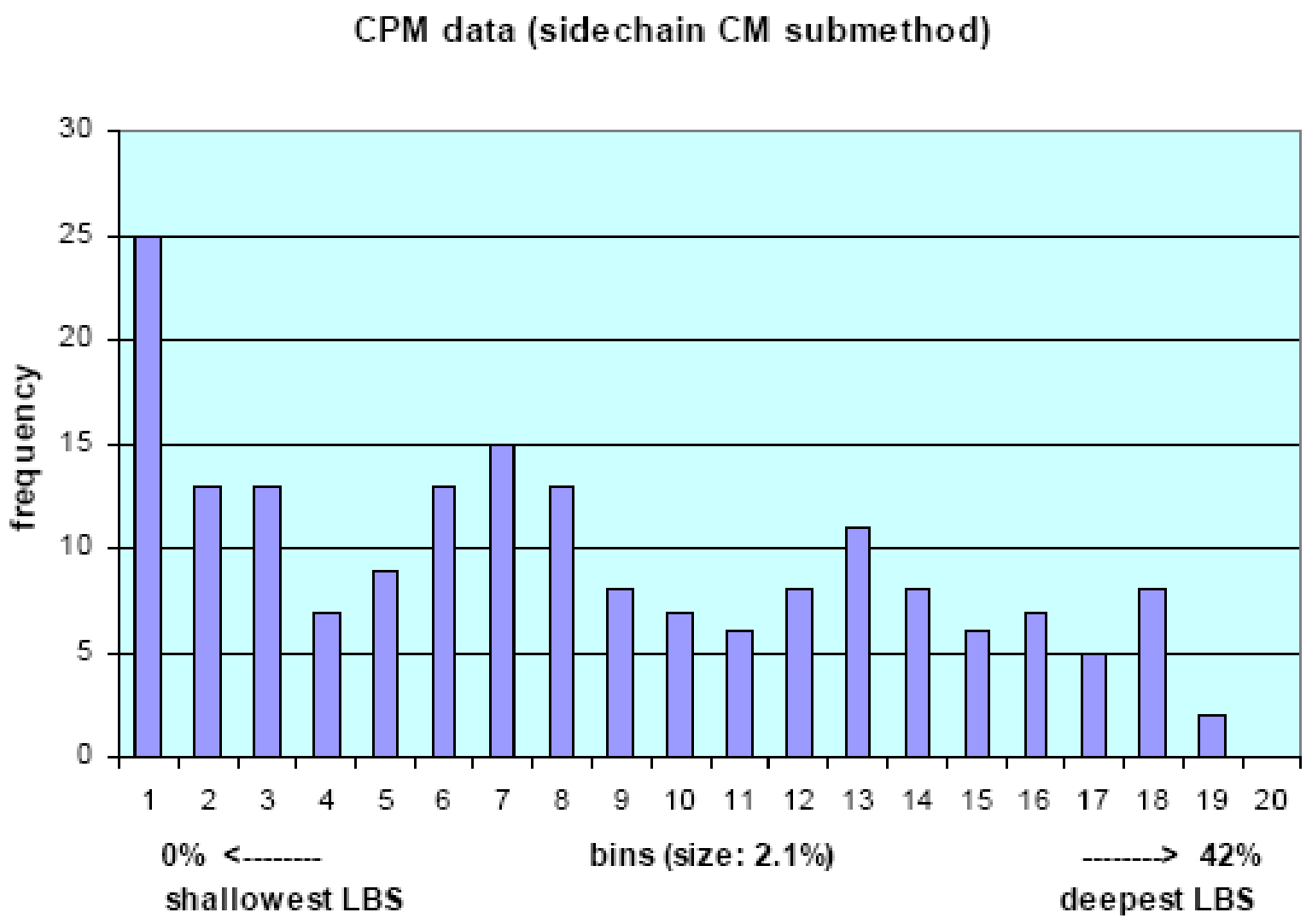

**Figure 3B.3**

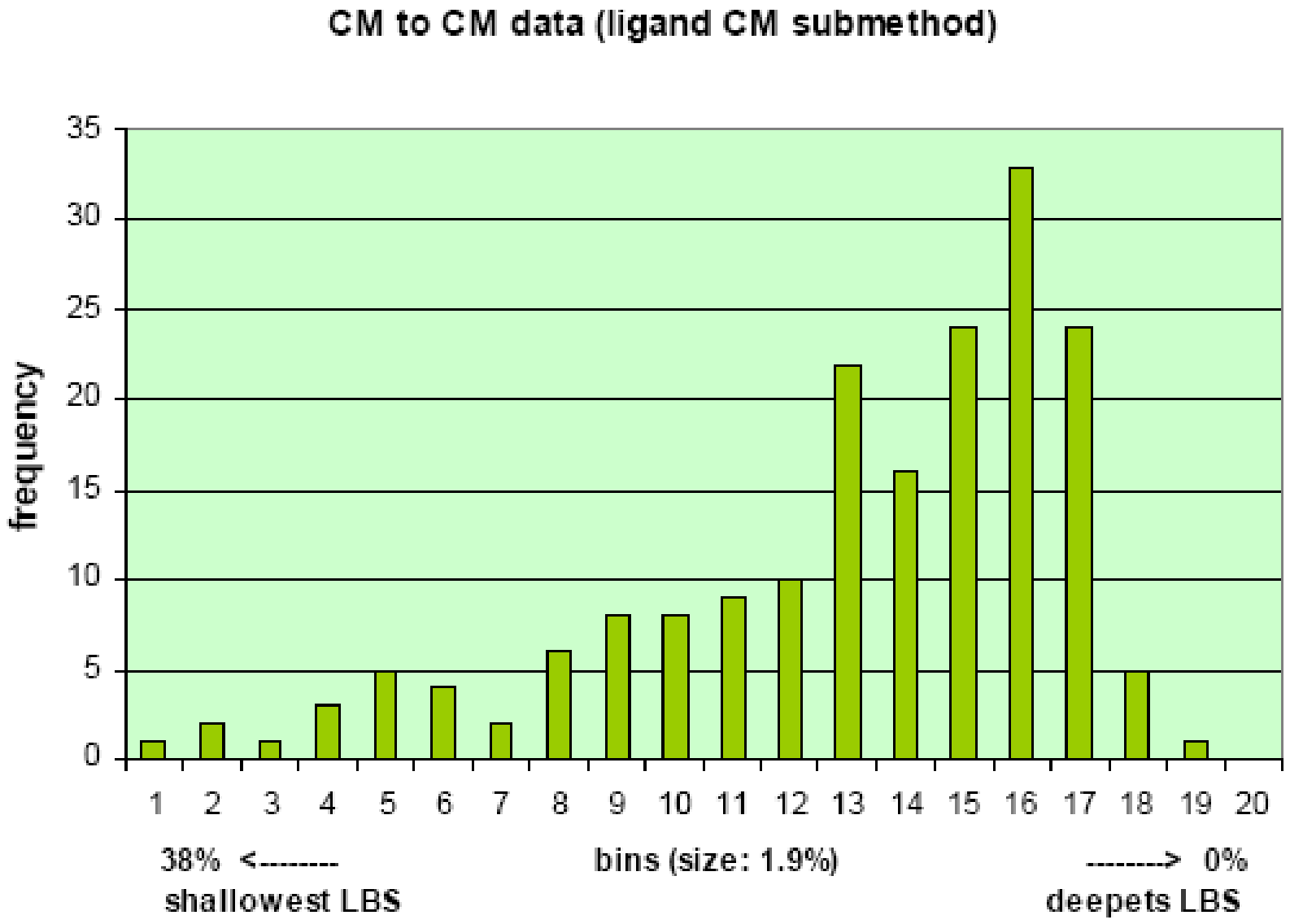

**Figure 3C.1**

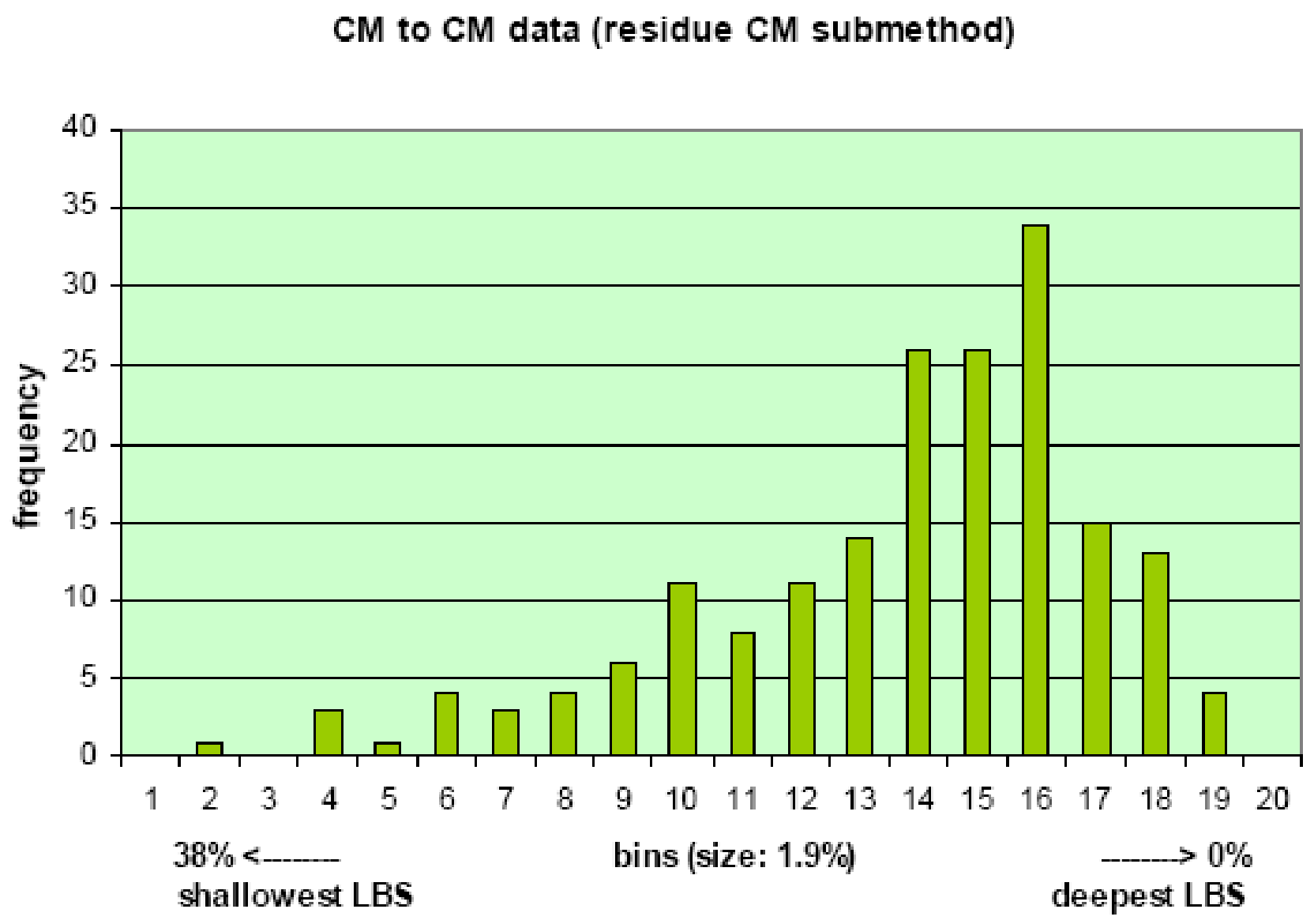

**Figure 3C.2**

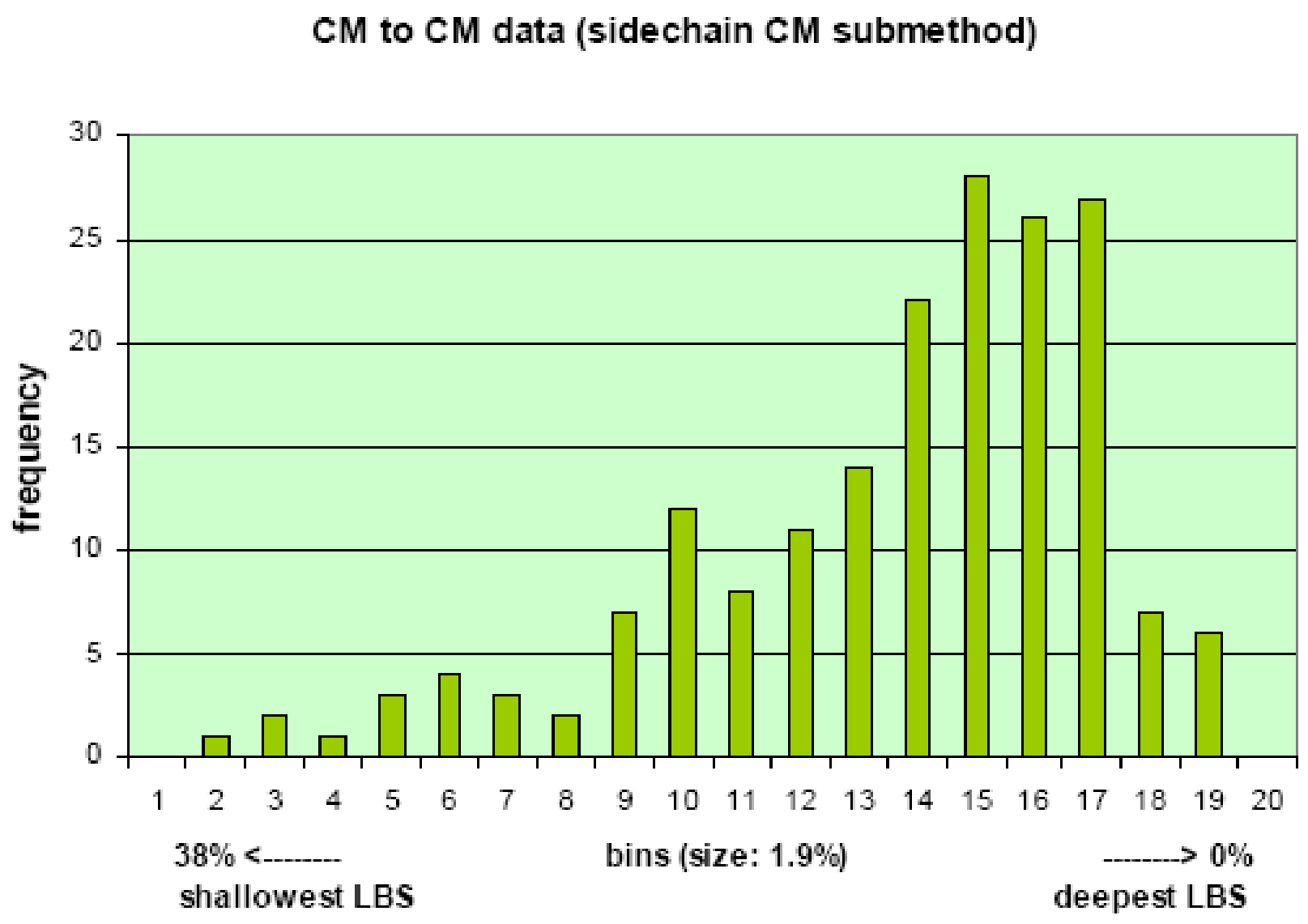

**Figure 3C.3**

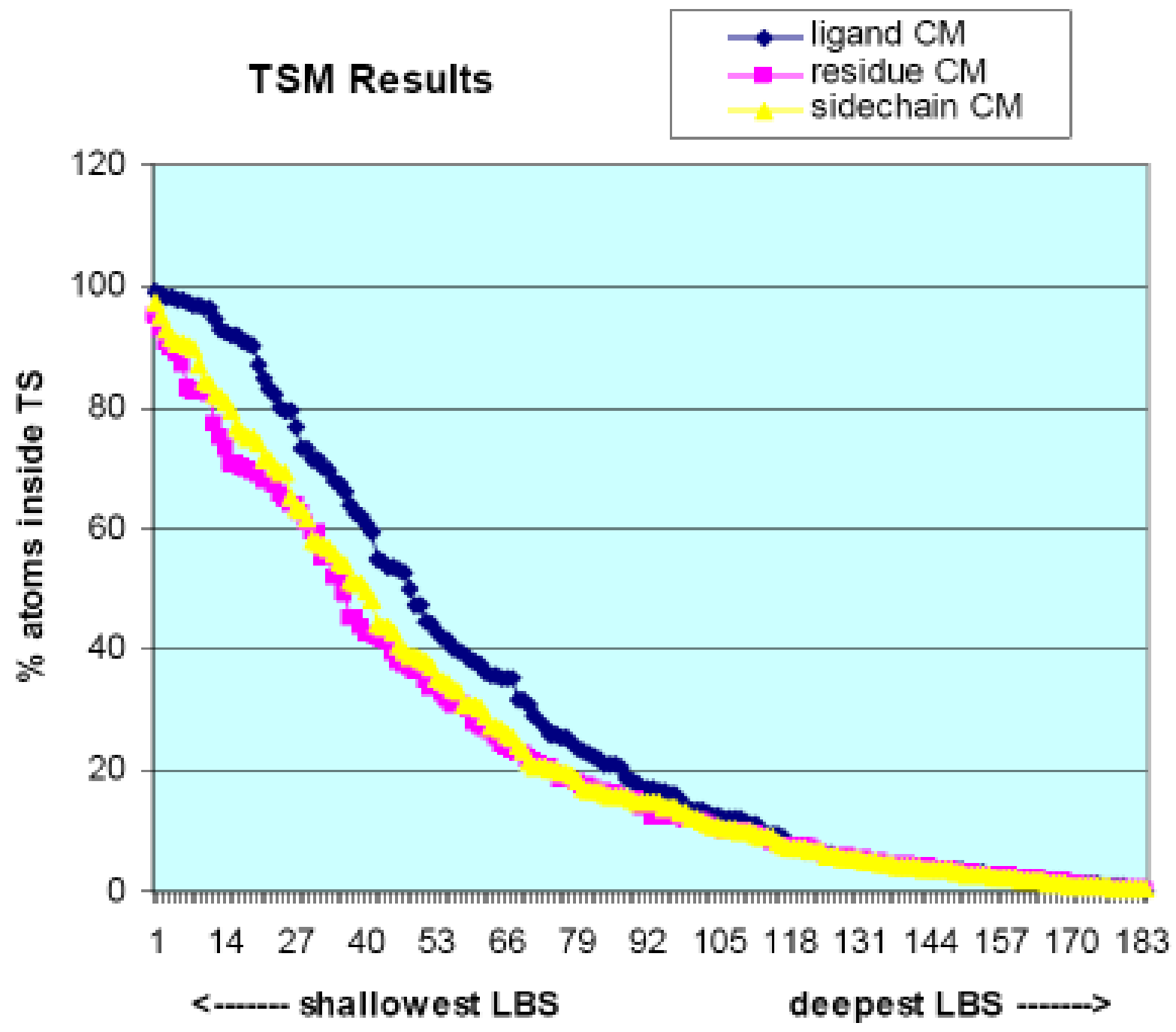

**Figure 4A**

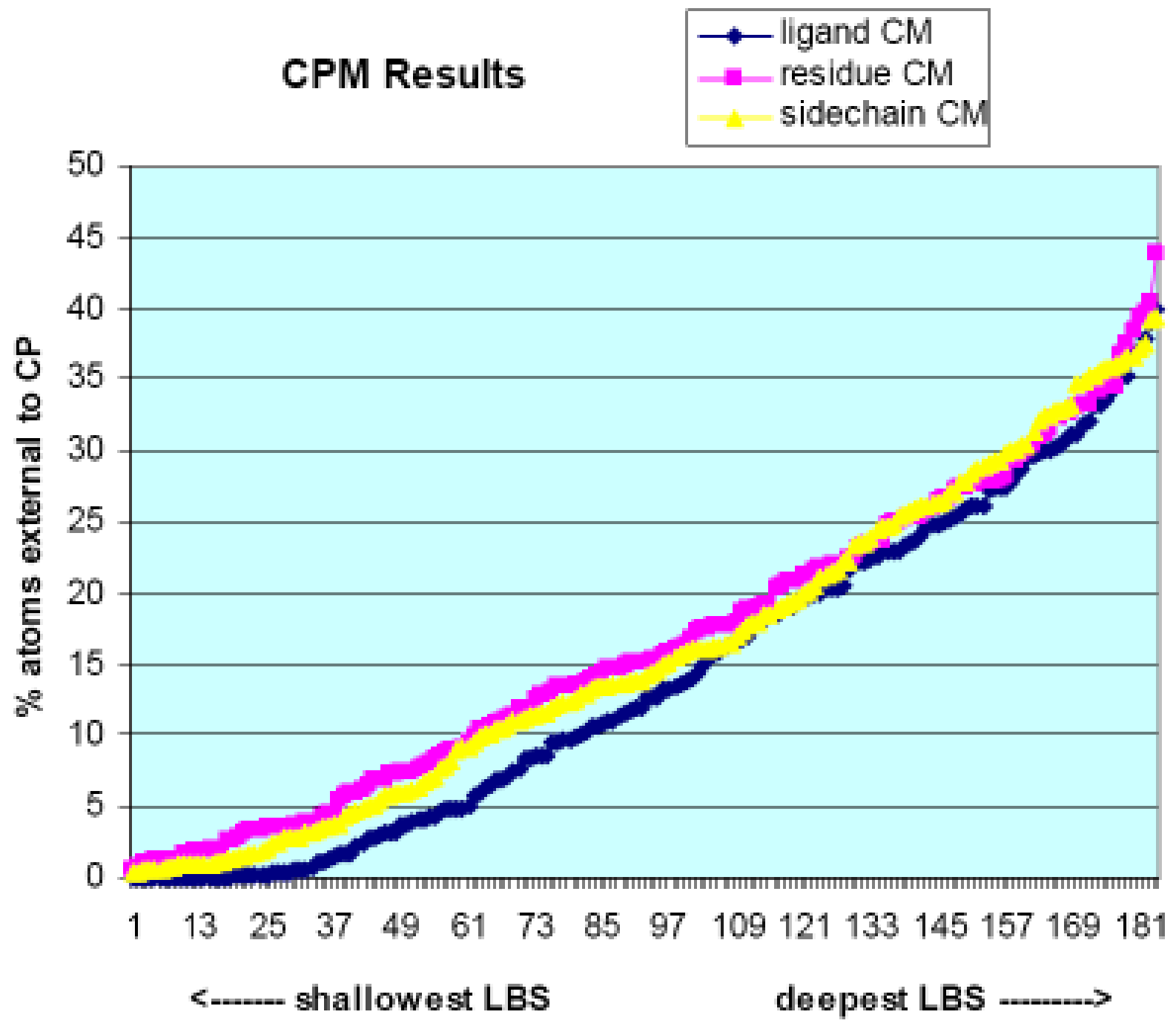

**Figure 4B**

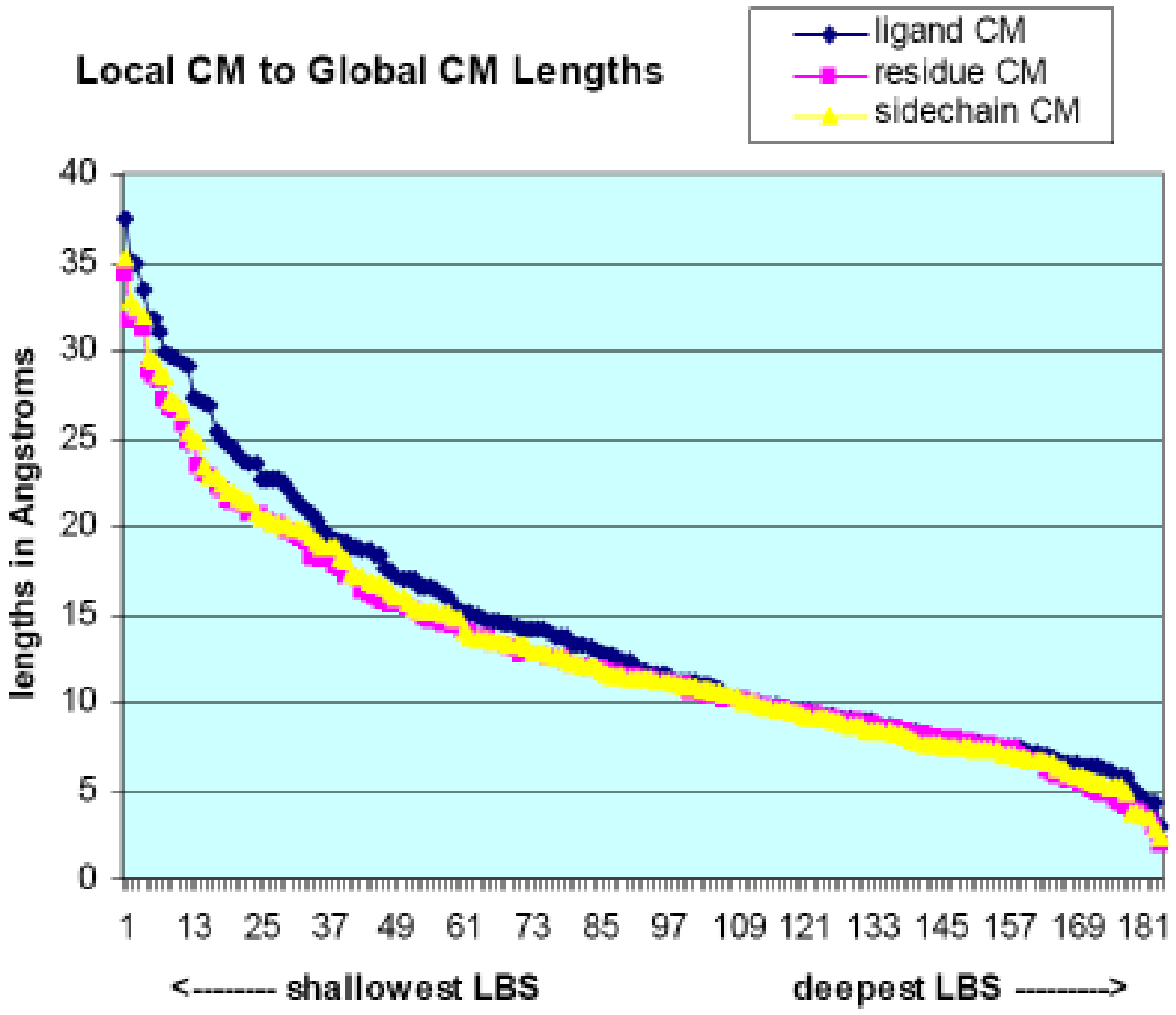

Figure 4C

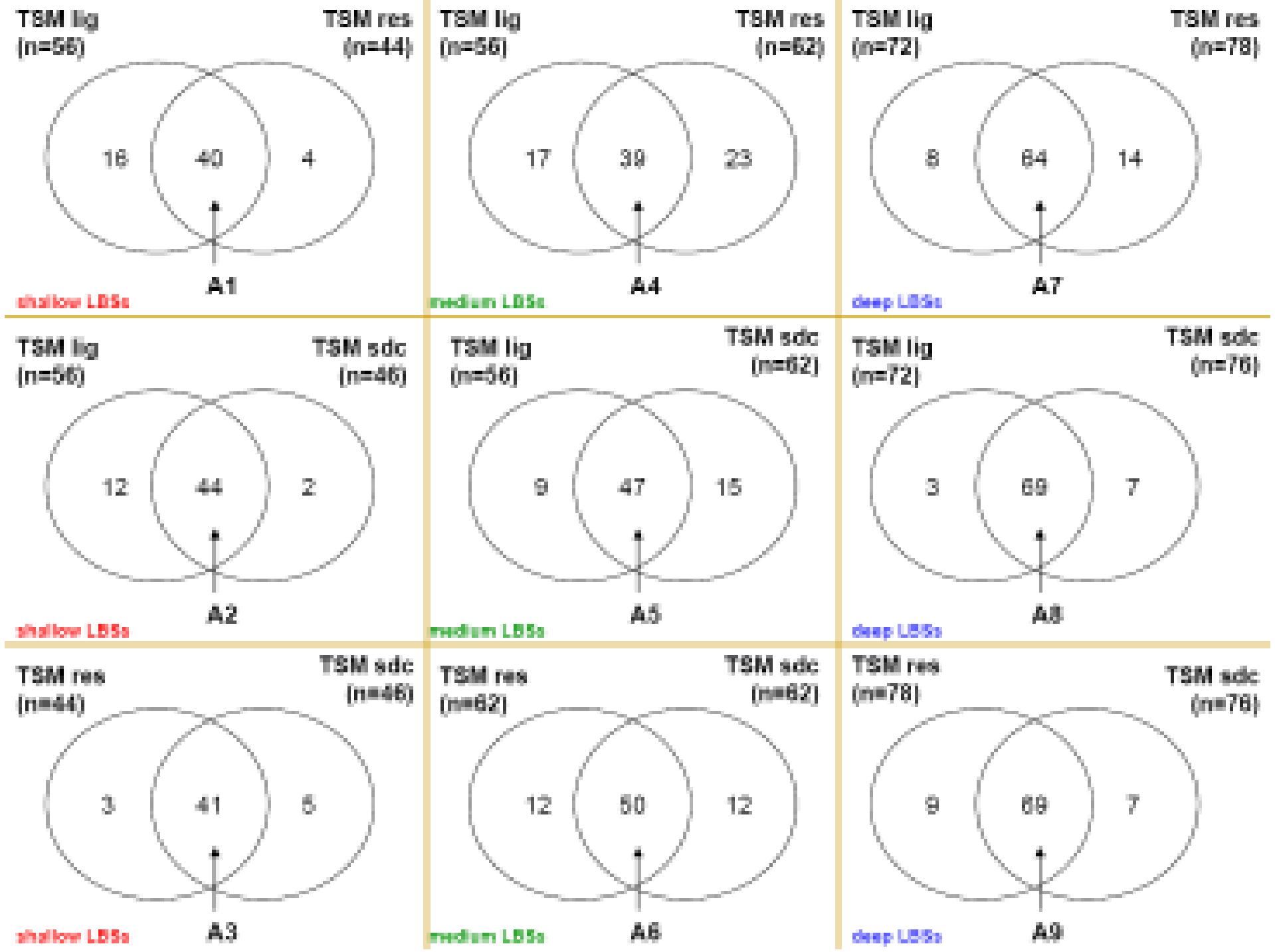

Figure 5A

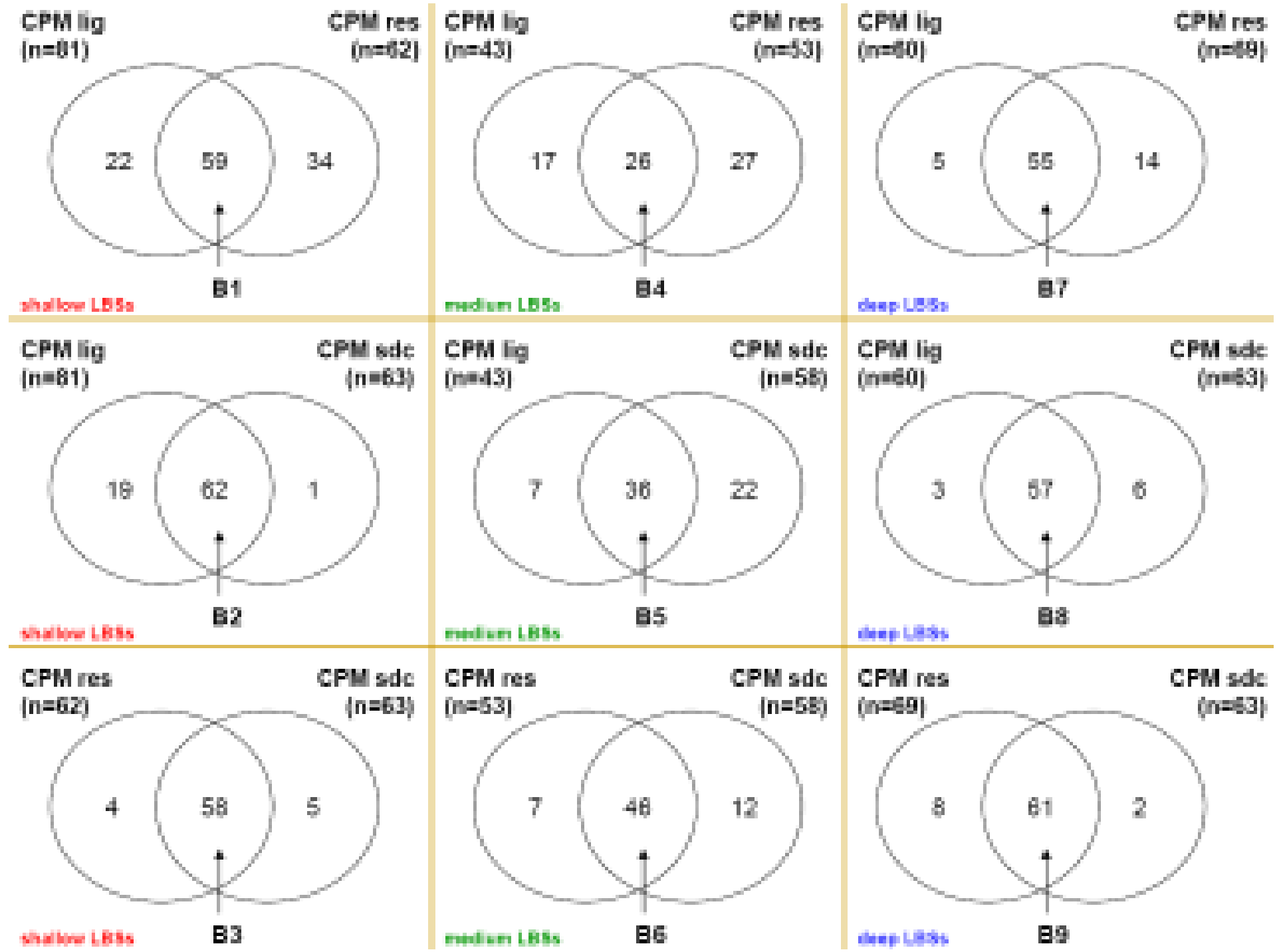

**Figure 5B**

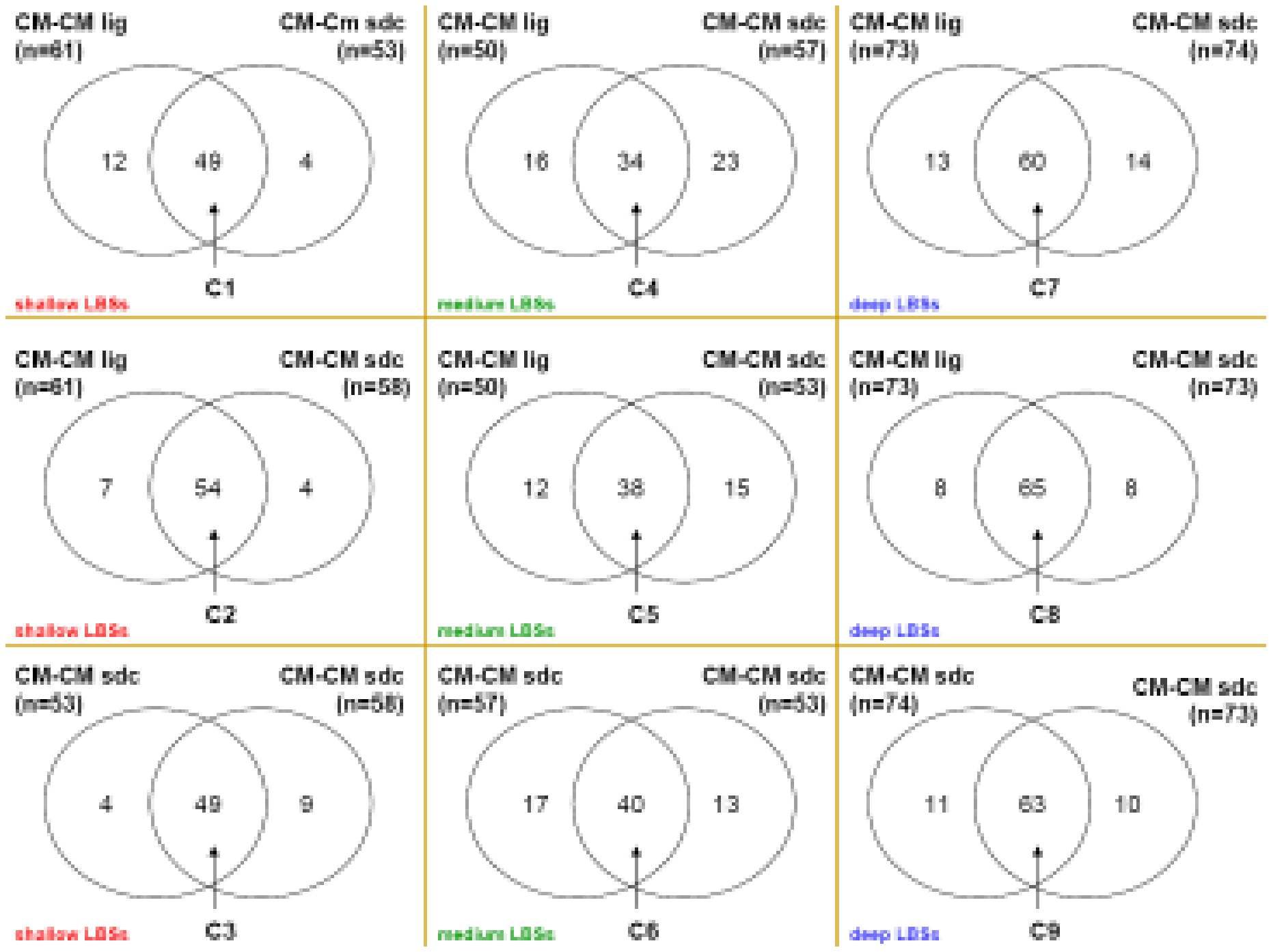

**Figure 5C**

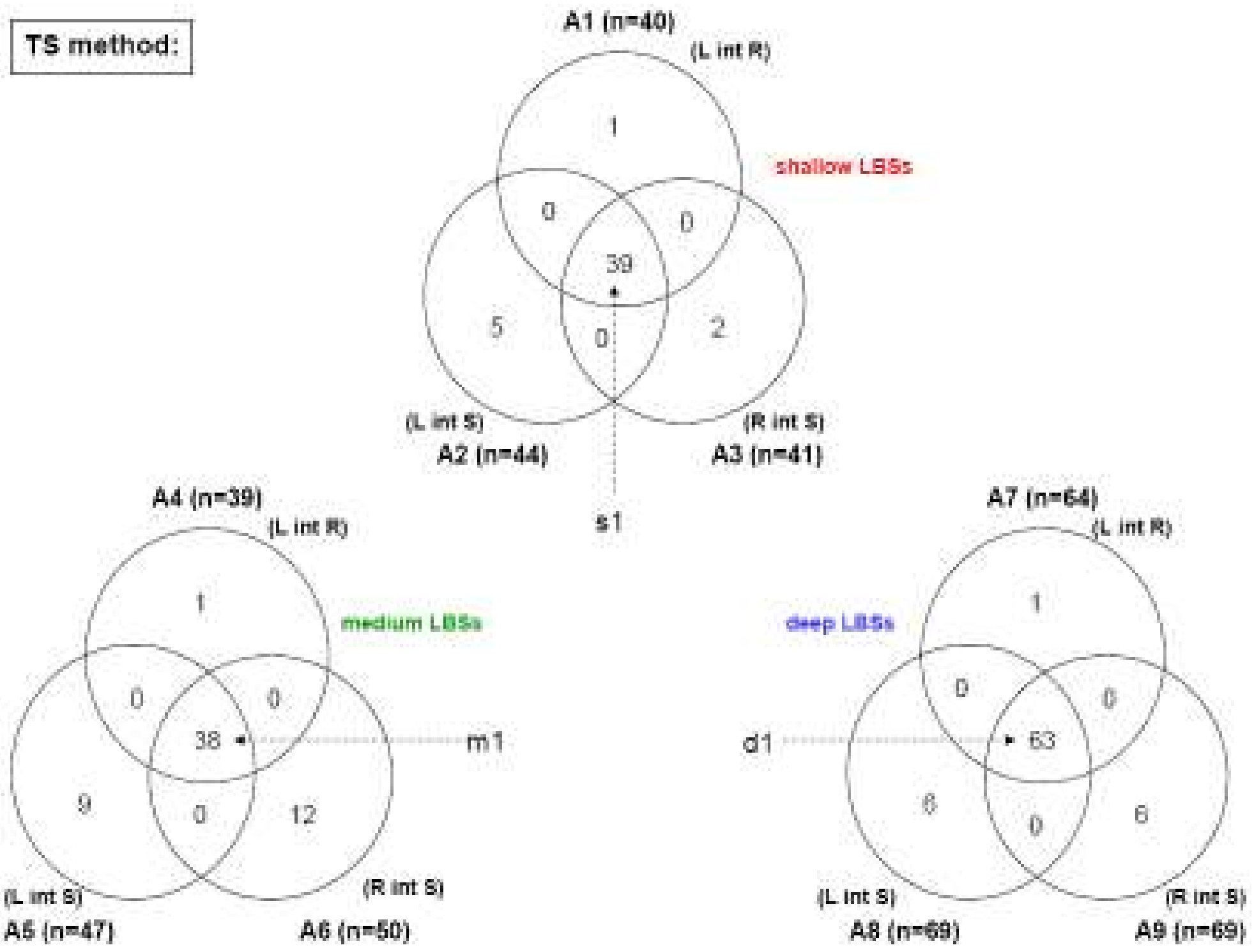

**Figure 6A**

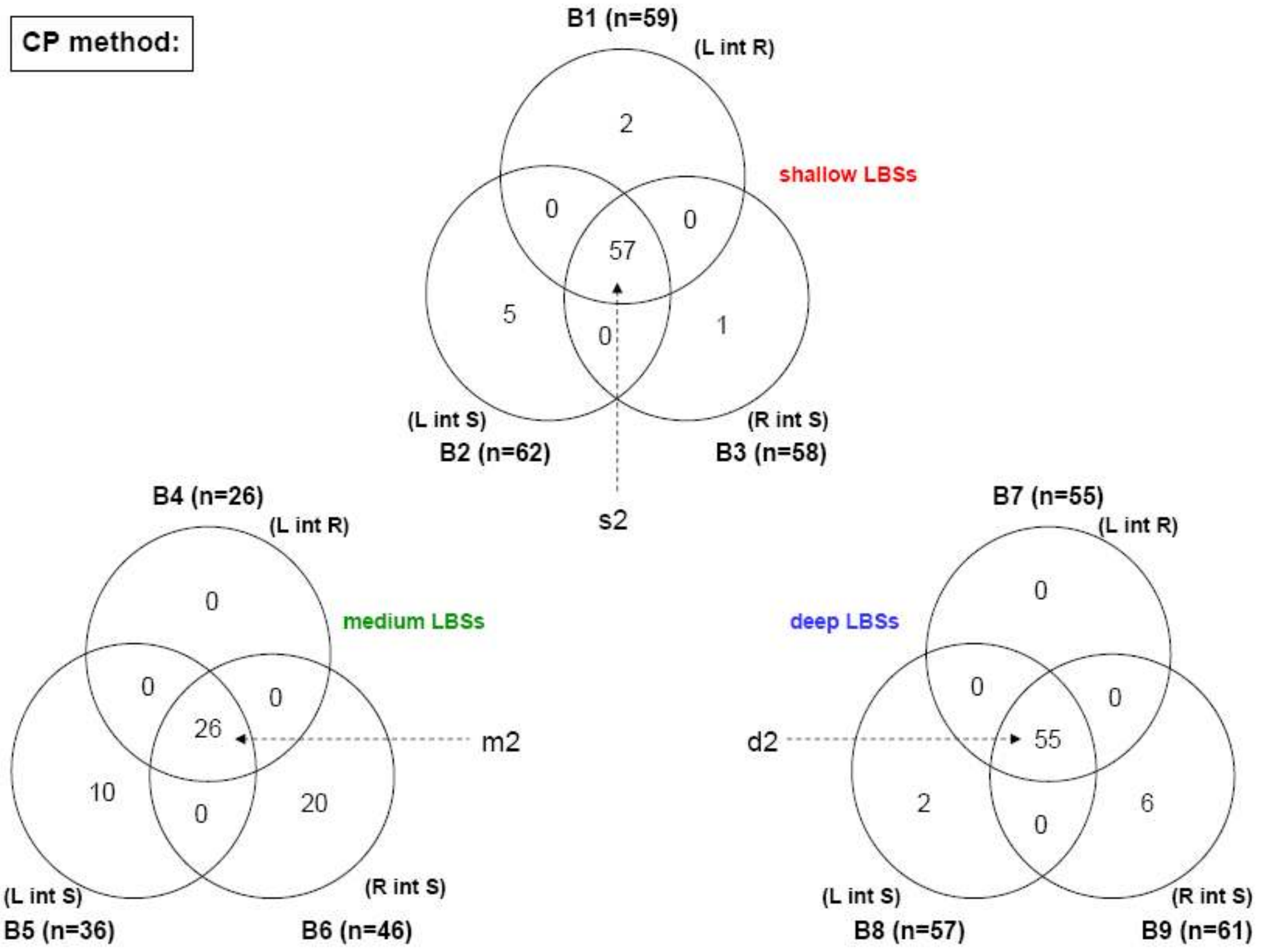

**Figure 6B**

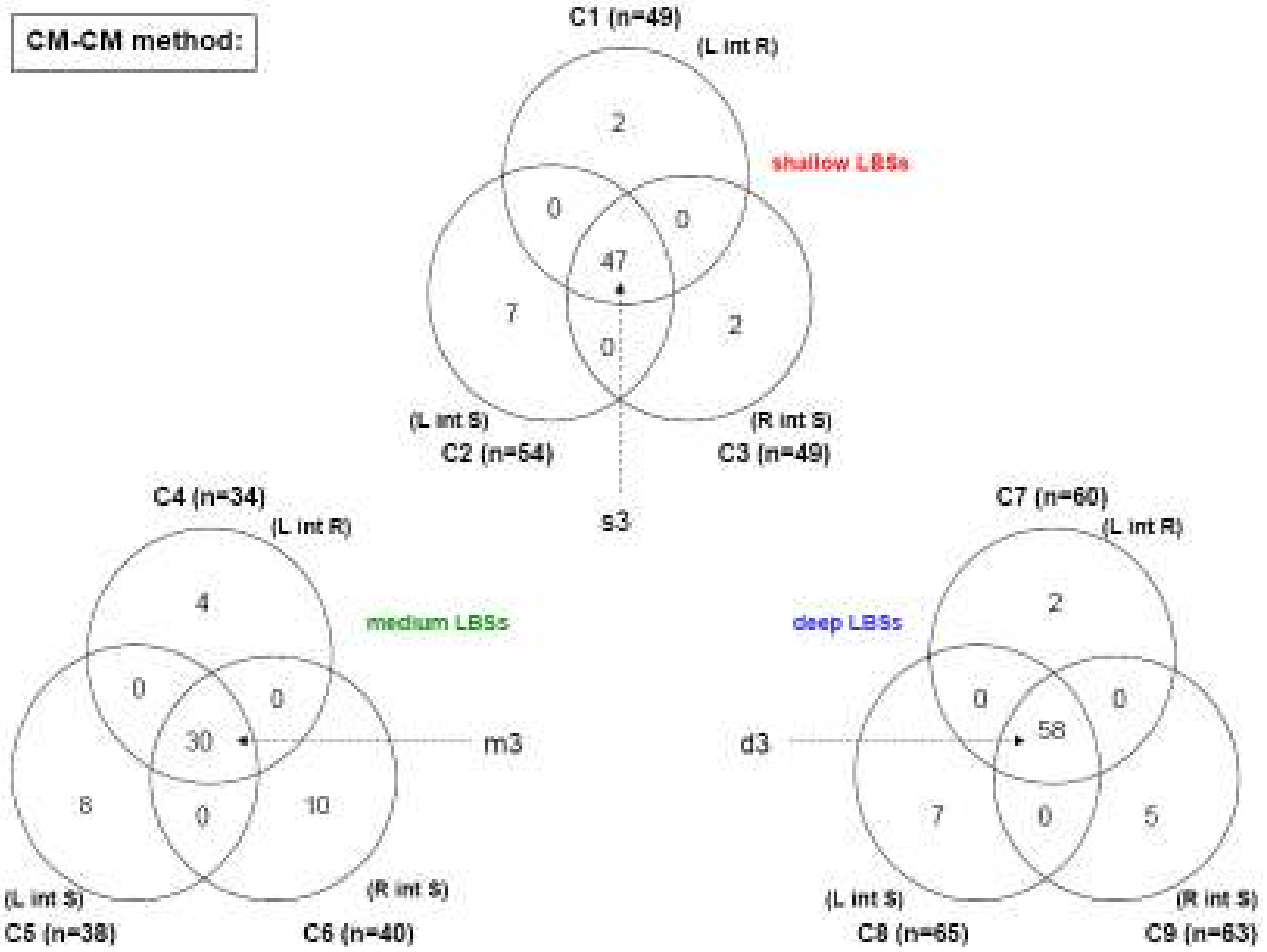

**Figure 6C**

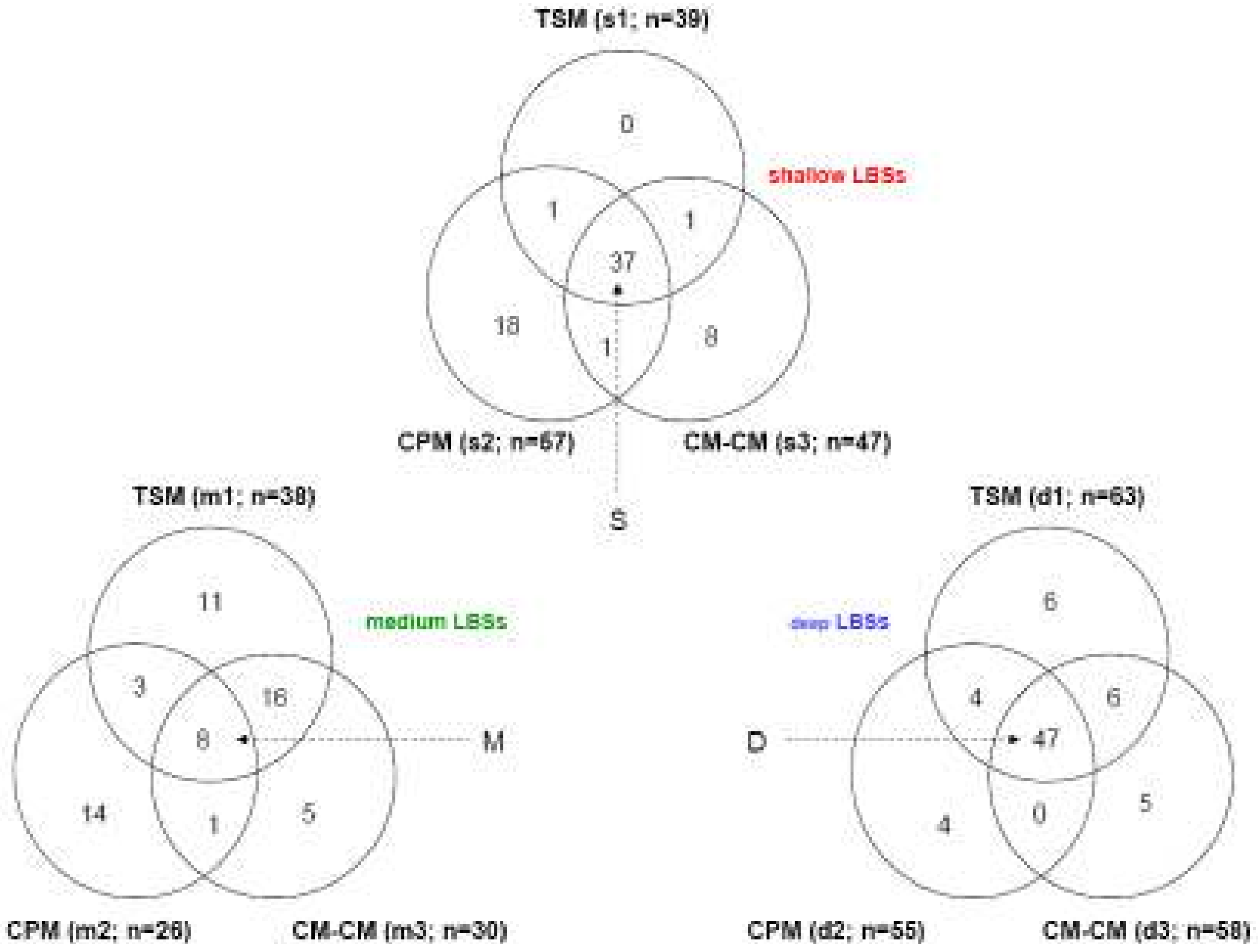

**Figure 7A  (CM-to-CM method included)**

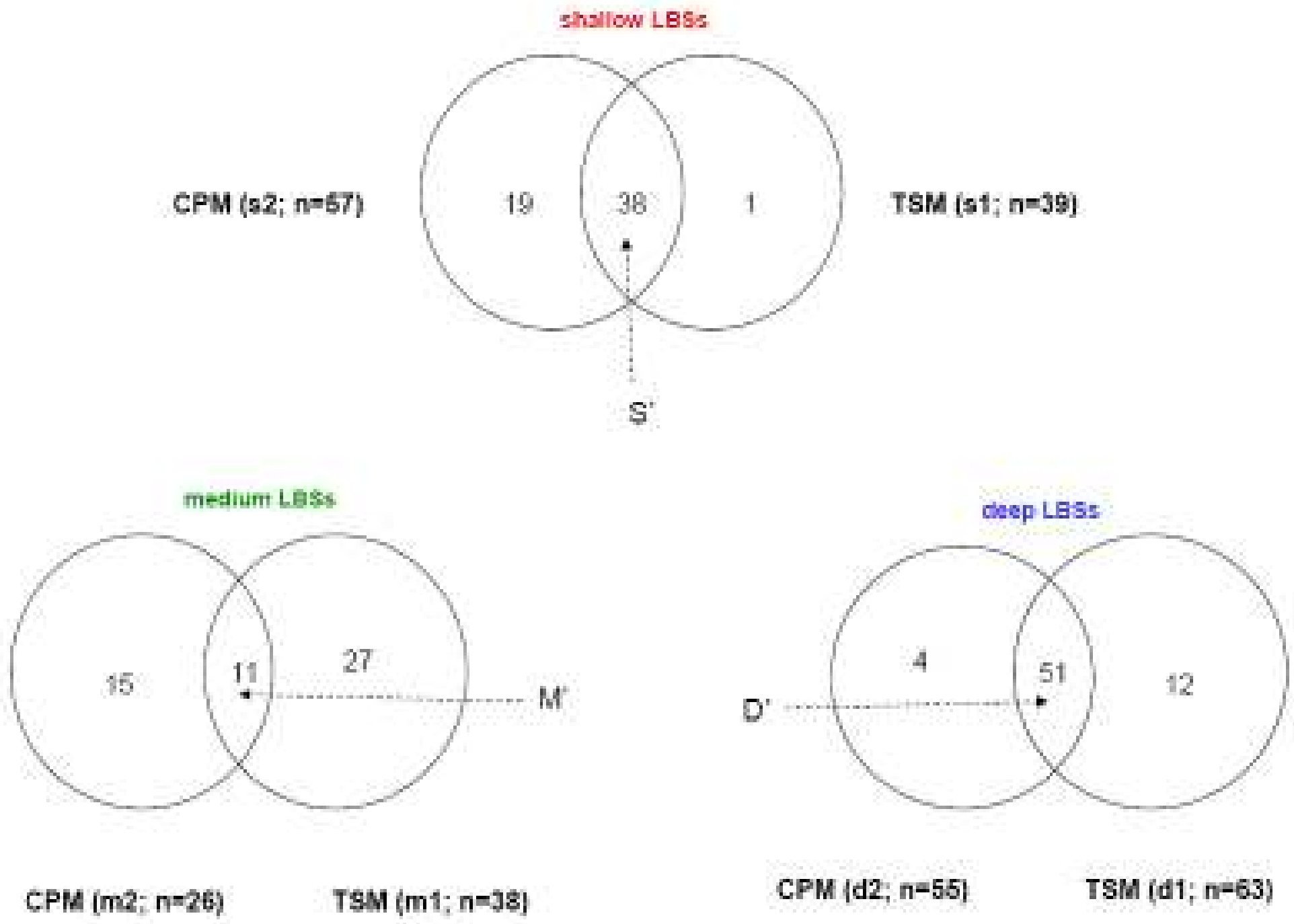

**Figure 7B (CM-to-CM method excluded)**

# TABLES:

Table 1 (part 1 of 4)

| Structure PDB ID | Shorthand designation | no. of domains | class | LIG # or LBS # | Ligand Name |
|---|---|---|---|---|---|
| 1ADS | Aa | 1 | 1 | 0 | NAP-350 |
| 9ICD | Ab | 1 | 1 | 0 | NAP-1 |
| 1IPD | Ac | 1 | 1 | 1 | SO4-346 |
| | | | | 2 | SO4-347 |
| 1GOX | Ad | 1 | 1 | 0 | FMN-1 |
| 1TDE | Ae | 2 | 1 | 0 | FAD-500 |
| 1OYB | Af | 1 | 1 | 1 | FMN-401 |
| | | | | 2 | HBA-402 |
| 2NPX | Ag | 3 | 1 | 1 | FAD-448 |
| | | | | 2 | NAD-818 |
| | | | | 3 | CYO-42 |
| 1CCA | Ah | 2 | 1 | 0 | HEM-1 |
| 1ARP | Ai | 2 | 1 | 1 | HEM-345 |
| | | | | 2 | NAG-348 |
| | | | | 3 | NAG-349 |
| 1PBE | Aj | 2 | 1 | 1 | FAD-395 |
| | | | | 2 | PHB-396 |
| 1HMY | Ak | 2 | 1 | 0 | SAM-328 |
| 3CLA | Al | 1 | 1 | 1 | CLM-221 |
| | | | | 2 | CO-222 |
| 1GPB | Am | 1 | 1 | 0 | PLP-999 |
| 1ULA | An | 1 | 1 | 1 | SO4-290 |
| | | | | 2 | SO4-291 |
| 1STO | Ao | 1 | 1 | 1 | OMP-216 |
| | | | | 2 | OH-214 |
| 2YHX | Ap | 3 | 1 | 0 | OTG-1 |
| 1PHP | Aq | 2 | 2 | 1 | ADP-395 |
| | | | | 2 | MG-1 |
| 1GKY | Ar | 2 | 1 | 1 | 5GP-187 |
| | | | | 2 | SO4-188 |
| 2CUT | As | 1 | 1 | 0 | DEP-401 |
| 1THG | At | 1 | 1 | 1 | NAG-990 |
| | | | | 2 | NAG-991 |
| | | | | 3 | NAG-994 |
| | | | | 4 | NAG-996 |
| 1RPA | Au | 1 | 1 | 1 | TAR-343 |
| | | | | 2 | NAG-344 |
| | | | | 3 | NAG-347 |
| 1RNH | Av | 1 | 3 | 1 | MSE-47 |
| | | | | 2 | MSE-50 |
| | | | | 3 | MSE-142 |
| | | | | 4 | SO4-156 |
| 1ONC | Aw | 1 | 2 | 0 | SO4-105 |
| 1FUT | Ax | 1 | 2 | 0 | 2GP-108 |
| 1ROB | Ay | 1 | 1 | 0 | C2P-126 |
| 1SNC | Az | 1 | 1 | 1 | PTP-143 |
| | | | | 2 | CA-142 |

**Table 1 (part 2 of 4)**

| | | | | | |
|---|---|---|---|---|---|
| 1CDG | Ba | 4 | 1 | 1 | MAL-688 |
| | | | | 2 | MAL-689 |
| | | | | 3 | MAL-690 |
| 1BYB | Bb | 1 | 1 | 1 | GLC-496 |
| | | | | 2 | GLC-497 |
| | | | | 3 | GLC-498 |
| | | | | 4 | GLC-499 |
| | | | | 5 | SO4-860 |
| 1XNB | Bc | 1 | 1 | 0 | SO4-191 |
| 2SIM | Bd | 1 | 2 | 0 | DAN-800 |
| 1BYH | Be | 1 | 1 | 1 | BUT-215 |
| | | | | 2 | GLC-216 |
| | | | | 3 | GLC-217 |
| 1FMP | Bf | 1 | 1 | 0 | FMP-301 |
| 1BLL | Bg | 2 | 1 | 1 | ZN-488 |
| | | | | 2 | ZN-489 |
| | | | | 3 | LEU(I-1) |
| | | | | 4 | FOR(I-2) |
| | | | | 5 | VAL(I-3) |
| | | | | 6 | VAL(I-4) |
| | | | | 7 | ASP(I-5) |
| 2CTC | Bh | 2 | 1 | 1 | LOF-309 |
| | | | | 2 | ZN-308 |
| 2GMT | Bi | 2 | 1 | 0 | HIN-247 |
| 1PPC | Bj | 2 | 1 | 1 | NAS(I-1) |
| | | | | 2 | APH(I-3) |
| | | | | 3 | PIP(I-4) |
| | | | | 4 | GLY(I-2) |
| 2ALP | Bk | 2 | 1 | 1 | SO4-1 |
| | | | | 2 | SO4-2 |
| 1ELA | Bl | 2 | 1 | 1 | TFA(B-256) |
| | | | | 2 | ISO(B-259) |
| | | | | 3 | SO4(B-290) |
| | | | | 4 | ACY(B-300) |
| | | | | 5 | LYS(B-257) |
| | | | | 6 | PRO(B-258) |
| 1HNE | Bm | 2 | 3 | 1 | ALM(I-5) |
| | | | | 2 | MSU(I-1) |
| | | | | 3 | ALA(I-2) |
| | | | | 4 | ALA(I-3) |
| | | | | 5 | PRO(I-4) |
| 1PEK | Bn | 2 | 1 | 1 | PRO(C-1) |
| | | | | 2 | ALA(C-2) |
| | | | | 3 | PRO(C-3) |
| | | | | 4 | PHE(C-4) |
| | | | | 5 | ALA(D-5) |
| | | | | 6 | ALA(D-6) |
| 3SGA | Bo | 2 | 1 | 1 | PHA(P-1) |
| | | | | 2 | ACE(P-5) |
| | | | | 3 | PRO(P-4) |
| | | | | 4 | ALA(P-3) |
| | | | | 5 | PRO(P-2) |

**Table 1 (part 3 of 4)**

| | | | | | |
|---|---|---|---|---|---|
| 1PIP | Bp | 2 | 1 | 1 | NIT(B-219) |
| | | | | 2 | SIN(B-213) |
| | | | | 3 | GLN(B-400) |
| | | | | 4 | VAL(B-401) |
| | | | | 5 | VAL(B-402) |
| | | | | 6 | ALA(B-403) |
| | | | | 7 | ALA(B-404) |
| 1SMR | Bq | 2 | 1 | 1 | LPL(B-6) |
| | | | | 2 | PIV(B-1) |
| | | | | 3 | HIS(B-2) |
| | | | | 4 | PRO(B-3) |
| | | | | 5 | PHE(B-4) |
| | | | | 6 | HIS(B-5) |
| | | | | 7 | TYR(B-7) |
| | | | | 8 | TYR(B-8) |
| | | | | 9 | SER(B-9) |
| 1PPI | Br | 2 | 1 | 1 | IVA(I-324) |
| | | | | 2 | PLE(I-327) |
| | | | | 3 | OPH(I-328) |
| | | | | 4 | MAN-329) |
| | | | | 5 | XYS-330) |
| | | | | 6 | SO4-36) |
| | | | | 7 | VAL(I-325) |
| | | | | 8 | VAL(I-326) |
| 3APR | Bs | 2 | 1 | 1 | FRD(I-5) |
| | | | | 2 | PRO(I-2) |
| | | | | 3 | PHE(I-3) |
| | | | | 4 | HIS(I-4) |
| 1EPM | Bt | 2 | 1 | 1 | PSA(I-1) |
| | | | | 2 | SO4-1 |
| | | | | 3 | SO4-2 |
| | | | | 4 | SO4-3 |
| | | | | 5 | THR(I-5) |
| | | | | 6 | PHE(I-4) |
| | | | | 7 | GLN(I-3) |
| | | | | 8 | ALA(I-2) |
| | | | | 9 | LEU(I-2') |
| | | | | 10 | ARG(I-3') |
| | | | | 11 | GLU(I-4') |
| 1MPP | Bu | 2 | 1 | 0 | SO4-501 |
| 1HYT | Bv | 2 | 1 | 1 | BZS-807 |
| | | | | 2 | DMS-806 |
| | | | | 3 | ZN-805 |
| 1IAG | Bw | 1 | 1 | 1 | SO4-260 |
| | | | | 2 | ZN-999 |
| 1ADD | Bx | 1 | 3 | 1 | IDA-353 |
| | | | | 2 | ZN-400 |
| 2DHC | By | 1 | 3 | 0 | DCE-600 |
| 1PII | Bz | 2 | 1 | 1 | PO4-453 |
| | | | | 2 | PO4-454 |
| 2DKB | Ca | 2 | 1 | 1 | PLP-272 |
| | | | | 2 | MES-434 |

**Table 1 (part 4 of 4)**

| | | | | | |
|---|---|---|---|---|---|
| 1CSH | Cb | 1 | 1 | 1 | AMX-700 |
| | | | | 2 | OAA-702 |
| 2POR | Cc | 1 | 1 | 1 | OTE-545 |
| | | | | 2 | OTE-546 |
| | | | | 3 | OTE-547 |
| | | | | 4 | OTE-548 |
| 1CIL | Cd | 1 | 2 | 1 | ETS-262 |
| | | | | 2 | ZN-261 |
| 8ACN | Ce | 3 | 3 | 1 | NIC-755 |
| | | | | 2 | FS4-999 |
| 5ENL | Cf | 2 | 1 | 1 | 2PG-442 |
| | | | | 2 | CA-438 |
| 1PDA | Cg | 3 | 1 | 1 | DPM-314 |
| | | | | 2 | ACY-315 |
| 1MNS | Ch | 2 | 1 | 1 | APG-166 |
| | | | | 2 | MG-360 |
| 3PGM | Ci | 1 | 1 | 1 | MP3-3 |
| | | | | 2 | SO2-2 |
| | | | | 3 | SO1-1 |
| 1BIB | Cj | 3 | 1 | 0 | BTN-500 |
| 2ABK | Ck | 2 | 1 | 0 | FS4-300 |
| 2ACK | Cl | 1 | 1 | 0 | EDR-999 |
| 2CND | Cm | 2 | 1 | 0 | FAD-271 |
| 2PGD | Cn | 2 | 2 | 1 | SO4-505 |
| | | | | 2 | SO4-507 |
| | | | | 3 | SO4-508 |
| 4BCL | Co | 1 | 1 | 1 | BCL-1 |
| | | | | 2 | BCL-2 |
| | | | | 3 | BCL-3 |
| | | | | 4 | BCL-4 |
| | | | | 5 | BCL-5 |
| | | | | 6 | BCL-6 |
| | | | | 7 | BCL-7 |

| Tangent Sphere Method | | | | Cutting Plane Method | | | | | CM-CM Method | | | | |
|---|---|---|---|---|---|---|---|---|---|---|---|---|---|
| shallow | medium | deep | | shallow | | medium | deep | | shallow | | medium | deep | |
| Ac.lbs1 | Ab.lbs0 | Aa.lbs0 | Bt.lbs1 | Ab.lbs0 | Bo.lbs2 | Al.lbs1 | Aa.lbs0 | Bt.lbs1 | Ab.lbs0 | Bm.lbs2 | An.lbs2 | Ad.lbs0 | Br.lbs2 |
| Ac.lbs2 | Ak.lbs0 | Ad.lbs0 | Bt.lbs7 | Ac.lbs1 | Bo.lbs3 | Ao.lbs1 | Ad.lbs0 | Bt.lbs7 | Ac.lbs1 | Bn.lbs1 | As.lbs0 | Ae.lbs0 | Br.lbs3 |
| Ai.lbs2 | An.lbs2 | Ae.lbs0 | Bt.lbs8 | Ac.lbs2 | Bo.lbs4 | Be.lbs2 | Ae.lbs0 | Bv.lbs1 | Ac.lbs2 | Bo.lbs2 | Au.lbs2 | Af.lbs1 | Br.lbs7 |
| Ai.lbs3 | Aq.lbs1 | Af.lbs1 | Bv.lbs1 | Ai.lbs2 | Bo.lbs5 | Bg.lbs1 | Af.lbs1 | Bv.lbs3 | Ai.lbs2 | Bp.lbs2 | Av.lbs4 | Af.lbs2 | Bs.lbs3 |
| Al.lbs2 | Au.lbs2 | Af.lbs2 | Bv.lbs3 | Ai.lbs3 | Bp.lbs1 | Bg.lbs4 | Af.lbs2 | Bx.lbs1 | Ai.lbs3 | Bp.lbs3 | Az.lbs1 | Ag.lbs2 | Bt.lbs1 |
| At.lbs1 | Ax.lbs0 | Ag.lbs1 | Bx.lbs1 | Al.lbs2 | Bp.lbs2 | Bg.lbs5 | Ag.lbs1 | Bx.lbs2 | Al.lbs2 | Bp.lbs4 | Be.lbs3 | Ag.lbs3 | Bt.lbs7 |
| At.lbs2 | Az.lbs1 | Ag.lbs2 | Bx.lbs2 | Ao.lbs2 | Bp.lbs3 | Bh.lbs1 | Ag.lbs2 | Bz.lbs1 | At.lbs1 | Bq.lbs9 | Bh.lbs1 | Ai.lbs1 | Bv.lbs1 |
| At.lbs3 | Bb.lbs4 | Ag.lbs3 | By.lbs0 | At.lbs1 | Bp.lbs4 | Bj.lbs2 | Ag.lbs3 | Ca.lbs1 | At.lbs2 | Br.lbs4 | Bi.lbs0 | Aj.lbs1 | Bv.lbs3 |
| At.lbs4 | Be.lbs3 | Ai.lbs1 | Bz.lbs1 | At.lbs2 | Bp.lbs5 | Bl.lbs1 | Ai.lbs1 | Ca.lbs2 | At.lbs3 | Br.lbs6 | Bj.lbs2 | Am.lbs0 | Bx.lbs1 |
| Au.lbs3 | Bg.lbs2 | Aj.lbs1 | Ca.lbs1 | At.lbs3 | Bp.lbs6 | Bl.lbs4 | Aj.lbs1 | Cb.lbs2 | At.lbs4 | Bt.lbs2 | Bj.lbs4 | An.lbs1 | Bx.lbs2 |
| Av.lbs3 | Bg.lbs3 | Am.lbs0 | Ca.lbs2 | At.lbs4 | Bp.lbs7 | Bm.lbs1 | Aj.lbs2 | Cd.lbs1 | Au.lbs3 | Bt.lbs3 | Bm.lbs4 | Ap.lbs0 | Bz.lbs1 |
| Ba.lbs1 | Bh.lbs1 | An.lbs1 | Cb.lbs2 | Au.lbs2 | Bq.lbs8 | Bn.lbs4 | Am.lbs0 | Cd.lbs2 | Av.lbs3 | Bt.lbs4 | Bm.lbs5 | Aq.lbs2 | Ca.lbs1 |
| Ba.lbs2 | Bi.lbs0 | Ap.lbs0 | Cd.lbs1 | Au.lbs3 | Bq.lbs9 | Bq.lbs3 | An.lbs1 | Ce.lbs1 | Ba.lbs1 | Bu.lbs0 | Bn.lbs2 | Ar.lbs1 | Ca.lbs2 |
| Ba.lbs3 | Bj.lbs2 | Aq.lbs2 | Cd.lbs2 | Av.lbs3 | Br.lbs4 | Br.lbs1 | Ap.lbs0 | Ce.lbs2 | Ba.lbs2 | Bv.lbs2 | Bn.lbs6 | Ar.lbs2 | Cb.lbs2 |
| Bb.lbs5 | Bj.lbs4 | Ar.lbs1 | Ce.lbs1 | Az.lbs1 | Br.lbs5 | Br.lbs3 | Aq.lbs1 | Cf.lbs1 | Ba.lbs3 | Bw.lbs1 | Bo.lbs4 | Au.lbs1 | Cd.lbs1 |
| Bc.lbs0 | Bk.lbs1 | Au.lbs1 | Ce.lbs2 | Ba.lbs1 | Br.lbs6 | Br.lbs8 | Aq.lbs2 | Cf.lbs2 | Bb.lbs5 | Bz.lbs2 | Bo.lbs5 | Av.lbs1 | Cd.lbs2 |
| Bg.lbs7 | Bl.lbs5 | Av.lbs1 | Cf.lbs1 | Ba.lbs2 | Bt.lbs2 | Bs.lbs2 | Ar.lbs1 | Cg.lbs1 | Bc.lbs0 | Cb.lbs1 | Bp.lbs6 | Av.lbs2 | Ce.lbs1 |
| Bm.lbs2 | Bm.lbs5 | Av.lbs2 | Cf.lbs2 | Ba.lbs3 | Bt.lbs3 | Bs.lbs4 | Au.lbs1 | Cg.lbs2 | Bg.lbs1 | Cc.lbs2 | Bp.lbs7 | Aw.lbs0 | Ce.lbs2 |
| Bn.lbs1 | Bn.lbs2 | Ay.lbs0 | Cg.lbs1 | Bb.lbs5 | Bt.lbs4 | Bt.lbs6 | Av.lbs1 | Ch.lbs1 | Bg.lbs2 | Cc.lbs3 | Bq.lbs3 | Ay.lbs0 | Cf.lbs1 |
| Bo.lbs2 | Bn.lbs3 | Bb.lbs1 | Cg.lbs2 | Bc.lbs0 | Bu.lbs0 | Bt.lbs8 | Av.lbs2 | Ch.lbs2 | Bg.lbs3 | Ck.lbs0 | Bq.lbs4 | Az.lbs2 | Cf.lbs2 |
| Bo.lbs3 | Bn.lbs6 | Bb.lbs2 | Ch.lbs1 | Be.lbs3 | Bv.lbs2 | Bt.lbs9 | Bb.lbs2 | Ci.lbs1 | Bg.lbs4 | Cn.lbs3 | Bq.lbs7 | Bb.lbs2 | Cg.lbs1 |
| Bp.lbs2 | Bo.lbs5 | Bb.lbs3 | Ch.lbs2 | Bg.lbs7 | Bw.lbs1 | Bw.lbs2 | Bb.lbs3 | Ci.lbs2 | Bg.lbs5 | Co.lbs1 | Bq.lbs8 | Bb.lbs3 | Cg.lbs2 |
| Bp.lbs3 | Bp.lbs6 | Be.lbs1 | Ci.lbs1 | Bj.lbs1 | Bz.lbs2 | Cb.lbs1 | Bf.lbs0 | Cl.lbs0 | Bg.lbs6 | Co.lbs2 | Br.lbs1 | Be.lbs1 | Ch.lbs1 |
| Bp.lbs4 | Bp.lbs7 | Bf.lbs0 | Ci.lbs2 | Bk.lbs2 | Cc.lbs2 | Cc.lbs1 | Bh.lbs2 | Cn.lbs2 | Bg.lbs7 | | Bs.lbs2 | Bf.lbs0 | Ci.lbs1 |
| Bq.lbs9 | Bq.lbs2 | Bh.lbs2 | Cj.lbs0 | Bl.lbs3 | Cc.lbs3 | Ci.lbs3 | Bq.lbs1 | Co.lbs4 | | | Bs.lbs4 | Bh.lbs2 | Ci.lbs2 |
| Br.lbs4 | Bq.lbs3 | Bq.lbs1 | Cl.lbs0 | Bl.lbs6 | Cc.lbs4 | Co.lbs2 | Br.lbs2 | Co.lbs5 | | | Bt.lbs5 | Bl.lbs4 | Cj.lbs0 |
| Br.lbs6 | Bq.lbs7 | Bq.lbs5 | Cn.lbs1 | Bm.lbs2 | Ck.lbs0 | | Br.lbs7 | Co.lbs6 | | | Bt.lbs6 | Bm.lbs1 | Cl.lbs0 |
| Bt.lbs2 | Bq.lbs8 | Br.lbs2 | Cn.lbs2 | Bm.lbs3 | Cn.lbs3 | | Bs.lbs1 | | | | Bt.lbs9 | Bq.lbs1 | Cn.lbs2 |
| Bt.lbs3 | Br.lbs1 | Br.lbs3 | Co.lbs5 | Bn.lbs1 | | | | | | | Bw.lbs2 | Bq.lbs5 | Co.lbs5 |
| Bt.lbs4 | Bs.lbs2 | Br.lbs7 | Co.lbs6 | | | | | | | | Cm.lbs0 | | |
| Bu.lbs0 | Bt.lbs5 | Bs.lbs3 | Co.lbs7 | | | | | | | | | | |
| Bv.lbs2 | Bt.lbs6 | Bs.lbs4 | | | | | | | | | | | |
| Bw.lbs1 | Bt.lbs9 | | | | | | | | | | | | |
| Bz.lbs2 | Bw.lbs2 | | | | | | | | | | | | |
| Cc.lbs2 | Cb.lbs1 | | | | | | | | | | | | |
| Cc.lbs3 | Cc.lbs1 | | | | | | | | | | | | |
| Ck.lbs0 | Cm.lbs0 | | | | | | | | | | | | |
| Cn.lbs3 | Co.lbs2 | | | | | | | | | | | | |
| Co.lbs1 | | | | | | | | | | | | | |

**Table2**

**Table 3A: Deep LBSs**

| STRUCTURE | T.S. Method | | | | C.P. Method | | | | CM-to-CM Method | | | |
|---|---|---|---|---|---|---|---|---|---|---|---|---|
| | lig | res | sdc | ave | lig | res | sdc | ave | lig | res | sdc | ave |
| Aa.lbs0 | 9.5105 | 6.3271 | 6.5261 | 7.45457 | 20.2547 | 21.9260 | 25.5074 | 22.5627 | 10.28162 | 9.04634 | 9.14596 | 9.49131 |
| Ad.lbs0 | 1.4079 | 0.2594 | 1.0745 | 0.91393 | 32.1230 | 34.4943 | 30.0482 | 32.2218 | 5.74836 | 3.76268 | 5.18857 | 4.89987 |
| Ae.lbs0 | 1.0204 | 3.8776 | 5.4762 | 3.45807 | 39.8639 | 34.1497 | 35.2041 | 36.4059 | 5.25637 | 8.20853 | 9.05430 | 7.50640 |
| Af.lbs1 | 1.3371 | 0.6171 | 0.3343 | 0.76283 | 29.8534 | 33.2219 | 35.8961 | 32.9905 | 5.89282 | 4.77545 | 3.65929 | 4.77585 |
| Af.lbs2 | 2.0057 | 2.4428 | 2.2885 | 2.24567 | 26.1764 | 26.5107 | 26.1507 | 26.2793 | 7.02114 | 7.35308 | 7.24344 | 7.20589 |
| Ag.lbs1 | 2.6075 | 7.3639 | 2.2063 | 4.05923 | 33.2092 | 25.7307 | 32.8940 | 30.6113 | 8.99809 | 12.47014 | 8.62754 | 10.03190 |
| Ag.lbs2 | 2.0057 | 3.3238 | 0.5444 | 1.95797 | 24.8424 | 25.2149 | 35.9312 | 28.6628 | 8.42256 | 9.70094 | 6.20313 | 8.10888 |
| Ag.lbs3 | 0.8309 | 2.1490 | 1.8052 | 1.59503 | 28.6533 | 27.6218 | 25.9885 | 27.4212 | 6.76621 | 8.49437 | 8.31799 | 7.85952 |
| Ai.lbs1 | 1.3387 | 0.7302 | 5.2333 | 2.43407 | 30.1014 | 30.9533 | 22.1907 | 27.7485 | 6.40161 | 5.44434 | 8.79111 | 6.87902 |
| Aj.lbs1 | 1.0975 | 2.5178 | 0.8070 | 1.47410 | 30.2776 | 27.1788 | 34.7321 | 30.7295 | 6.36432 | 7.77716 | 5.88053 | 6.67400 |
| Am.lbs0 | 0.8071 | 1.7486 | 1.2106 | 1.25543 | 39.0823 | 37.4682 | 39.2617 | 38.6041 | 6.52328 | 8.84403 | 7.43424 | 7.60052 |
| An.lbs1 | 4.3401 | 4.0301 | 3.7201 | 4.03010 | 23.4278 | 21.9221 | 24.7564 | 23.3688 | 7.82188 | 7.48854 | 7.27156 | 7.52733 |
| Ap.lbs0 | 2.2796 | 0.6079 | 1.0030 | 1.29683 | 27.3860 | 33.2827 | 34.8328 | 31.8338 | 7.47539 | 4.51663 | 5.34812 | 5.78005 |
| Aq.lbs2 | 4.6210 | 2.5931 | 3.6569 | 3.62367 | 25.9641 | 31.5160 | 27.8590 | 28.4464 | 9.72327 | 7.86723 | 8.93589 | 8.84213 |
| Ar.lbs1 | 2.8986 | 4.4859 | 4.3478 | 3.91077 | 31.1249 | 25.3968 | 26.2940 | 27.6052 | 6.36456 | 7.57155 | 7.43626 | 7.12412 |
| Au.lbs1 | 6.2634 | 5.2255 | 3.5075 | 4.99880 | 24.8032 | 26.5927 | 27.8096 | 26.4018 | 9.01569 | 8.48317 | 7.52633 | 8.34173 |
| Av.lbs1 | 6.6836 | 8.4602 | 2.7073 | 5.95037 | 22.1658 | 22.5042 | 29.1878 | 24.6193 | 7.76219 | 8.21716 | 5.73234 | 7.23723 |
| Av.lbs2 | 2.7919 | 1.1844 | 0.4230 | 1.46643 | 23.0118 | 28.2572 | 36.2098 | 29.1596 | 5.77724 | 4.04717 | 2.80917 | 4.21119 |
| Bb.lbs2 | 1.9873 | 0.3057 | 1.7070 | 1.33333 | 25.0955 | 38.4968 | 30.4713 | 31.3545 | 7.25613 | 3.90801 | 6.66775 | 5.94396 |
| Bb.lbs3 | 3.2611 | 3.2611 | 1.7580 | 2.76007 | 23.2357 | 25.4268 | 28.3057 | 25.6561 | 8.39079 | 8.33648 | 6.79967 | 7.84231 |
| Bf.lbs0 | 4.0208 | 0.6149 | 2.3652 | 2.33363 | 22.2327 | 34.5790 | 24.4560 | 27.0892 | 7.39599 | 4.26679 | 5.90191 | 5.85490 |
| Bh.lbs2 | 7.0434 | 4.8321 | 5.9787 | 5.95140 | 22.5225 | 25.0205 | 24.5700 | 24.0377 | 9.25730 | 8.09097 | 8.61611 | 8.65479 |
| Bq.lbs1 | 2.2700 | 1.9116 | 2.1505 | 2.11070 | 25.6472 | 27.8773 | 25.3286 | 26.2844 | 6.58757 | 6.10394 | 6.51448 | 6.40200 |
| Br.lbs2 | 2.4473 | 2.4051 | 1.2658 | 2.03940 | 27.1730 | 27.8461 | 33.0380 | 29.3530 | 6.55463 | 6.49278 | 4.89134 | 5.97958 |
| Br.lbs7 | 6.3713 | 5.7384 | 3.5021 | 5.20393 | 21.6456 | 22.0253 | 25.6540 | 23.1083 | 9.07211 | 8.79786 | 7.60733 | 8.49243 |
| Bt.lbs1 | 2.9720 | 1.5906 | 3.4324 | 2.66500 | 23.6919 | 29.2591 | 21.2641 | 24.7384 | 7.13280 | 5.43769 | 7.53525 | 6.70191 |
| Bt.lbs7 | 5.9021 | 2.6789 | 2.5952 | 3.72540 | 20.3851 | 27.5848 | 29.3428 | 25.7709 | 8.98414 | 6.91482 | 6.60925 | 7.50274 |
| Bv.lbs1 | 5.7014 | 5.2912 | 4.7170 | 5.23653 | 20.5906 | 20.8368 | 23.3798 | 21.6024 | 8.50683 | 8.21531 | 7.84684 | 8.18966 |
| Bv.lbs3 | 4.1838 | 2.2149 | 4.0607 | 3.48647 | 22.3134 | 28.0148 | 23.4208 | 24.5830 | 7.45971 | 5.76439 | 7.34854 | 6.85755 |
| Bx.lbs1 | 3.7966 | 5.9456 | 2.4713 | 4.07117 | 27.5072 | 25.0000 | 30.3367 | 27.6146 | 7.71042 | 9.08760 | 6.76607 | 7.85470 |
| Bx.lbs2 | 0.9670 | 0.2865 | 0.5731 | 0.60887 | 35.2436 | 40.4370 | 37.5716 | 37.7507 | 4.39819 | 2.90809 | 3.76015 | 3.68881 |
| Bz.lbs1 | 4.9092 | 4.9943 | 3.3201 | 4.40787 | 34.0806 | 34.0522 | 36.4926 | 34.8751 | 9.43109 | 9.46442 | 8.34145 | 9.07899 |
| Ca.lbs1 | 1.2915 | 0.5843 | 0.7688 | 0.88153 | 31.9803 | 39.3296 | 35.7319 | 35.6806 | 6.26963 | 5.01368 | 5.39555 | 5.55962 |
| Ca.lbs2 | 3.6593 | 0.7688 | 1.6913 | 2.03980 | 21.9557 | 30.3813 | 23.9545 | 25.4305 | 8.53387 | 5.59405 | 6.82285 | 6.98359 |
| Cb.lbs2 | 3.0964 | 1.7104 | 2.0643 | 2.29037 | 28.3102 | 32.2029 | 29.8437 | 30.1189 | 7.74503 | 6.52315 | 6.73587 | 7.00135 |
| Cd.lbs1 | 5.0490 | 2.5000 | 1.5196 | 3.02287 | 22.9902 | 25.9314 | 29.3627 | 26.0948 | 7.71650 | 6.02835 | 5.25182 | 6.33222 |
| Cd.lbs2 | 0.3922 | 0.1471 | 0.2451 | 0.26147 | 36.8137 | 43.7745 | 39.4118 | 40.0000 | 3.02225 | 1.90705 | 2.35654 | 2.42861 |
| Ce.lbs1 | 0.4818 | 0.4129 | 0.6710 | 0.52190 | 37.8871 | 39.6937 | 37.0440 | 38.2083 | 4.94249 | 4.78547 | 5.51390 | 5.08062 |
| Ce.lbs2 | 1.1700 | 1.5485 | 1.0840 | 1.26750 | 33.2416 | 34.6008 | 35.1858 | 34.3427 | 7.15173 | 7.85439 | 6.95883 | 7.32165 |
| Cf.lbs1 | 1.6418 | 0.3040 | 1.0033 | 0.98303 | 30.7692 | 36.7893 | 32.5023 | 33.3536 | 6.17896 | 3.82368 | 5.29381 | 5.09882 |
| Cf.lbs2 | 2.7668 | 2.5236 | 2.7364 | 2.67560 | 30.0091 | 32.2894 | 31.2861 | 31.1949 | 7.66465 | 7.38576 | 7.58245 | 7.54429 |
| Cg.lbs1 | 3.6080 | 7.4388 | 4.4543 | 5.16703 | 24.0535 | 21.9154 | 21.4699 | 22.4796 | 7.89290 | 9.95957 | 8.60208 | 8.81818 |
| Cg.lbs2 | 3.3853 | 2.8953 | 4.0535 | 3.44470 | 26.1470 | 30.0668 | 26.5033 | 27.5724 | 7.76433 | 7.26717 | 8.37459 | 7.80203 |
| Ch.lbs1 | 3.5582 | 4.2624 | 3.5582 | 3.79293 | 29.6887 | 31.1712 | 32.1720 | 31.0106 | 7.74771 | 8.30537 | 7.75728 | 7.93679 |
| Ch.lbs2 | 5.0037 | 9.3773 | 5.9674 | 6.78280 | 30.4299 | 27.8354 | 30.0964 | 29.4539 | 8.68576 | 10.66112 | 9.13461 | 9.49383 |
| Ci.lbs1 | 3.7643 | 5.0736 | 3.8734 | 4.23710 | 26.1866 | 23.2406 | 24.6590 | 24.6954 | 6.83944 | 7.64921 | 6.97797 | 7.15554 |
| Ci.lbs2 | 1.2002 | 3.6552 | 0.4910 | 1.78213 | 34.8063 | 26.6776 | 36.5521 | 32.6787 | 4.56176 | 6.72447 | 3.38018 | 4.88880 |
| Cl.lbs0 | 2.3637 | 1.9296 | 1.4231 | 1.90547 | 27.4964 | 32.5856 | 28.9918 | 29.6913 | 7.44867 | 6.97387 | 6.32670 | 6.91641 |
| Cn.lbs2 | 4.0219 | 3.5021 | 2.6539 | 3.39263 | 29.4938 | 33.0506 | 33.1874 | 31.9106 | 9.04226 | 8.51375 | 7.64118 | 8.39906 |
| Co.lbs5 | 0.0000 | 2.0956 | 0.4412 | 0.84560 | 37.0956 | 23.6765 | 32.4265 | 31.0662 | 4.25437 | 9.73766 | 7.64564 | 7.21256 |
| Co.lbs6 | 0.6618 | 4.5588 | 4.5221 | 3.24757 | 33.4926 | 27.3897 | 26.0662 | 28.9828 | 8.18778 | 11.55079 | 11.54632 | 10.42830 |

**Table 3B: Medium LBSs**

| STRUCTURE | T.S. Method | | | | C.P. Method | | | | CM-to-CM Method | | | |
|---|---|---|---|---|---|---|---|---|---|---|---|---|
| | lig | res | sdc | ave | lig | res | sdc | ave | lig | res | sdc | ave |
| Bh.lbs1 | 11.8755 | 17.4857 | 14.6192 | 14.6601 | 17.0352 | 13.5135 | 14.3735 | 14.9741 | 11.13238 | 12.69575 | 12.08921 | 11.9724 |
| Bj.lbs2 | 16.8201 | 23.5727 | 13.9963 | 18.1297 | 14.9171 | 13.3824 | 18.9687 | 15.7561 | 10.70033 | 12.09912 | 10.00316 | 10.9342 |
| Bq.lbs3 | 22.5010 | 11.4695 | 18.1203 | 17.3636 | 10.5934 | 14.7352 | 11.4695 | 12.2660 | 14.20903 | 11.40287 | 13.25944 | 12.9571 |
| Br.lbs1 | 16.2447 | 16.4979 | 14.7257 | 15.8228 | 14.4304 | 17.5105 | 16.4979 | 16.1463 | 12.38781 | 12.48107 | 12.06181 | 12.3102 |
| Bs.lbs2 | 17.0459 | 11.5019 | 11.9983 | 13.5154 | 12.6189 | 18.8664 | 17.9975 | 16.4943 | 12.80356 | 11.17469 | 11.34710 | 11.7751 |
| Bt.lbs6 | 17.6643 | 18.3759 | 19.9665 | 18.6689 | 13.5622 | 17.7062 | 16.2411 | 15.8365 | 12.93160 | 13.06783 | 13.44475 | 13.1481 |
| Bt.lbs9 | 12.1808 | 11.4274 | 10.0879 | 11.2320 | 10.7576 | 15.8644 | 16.4923 | 14.3714 | 11.49769 | 11.28031 | 10.79240 | 11.1901 |
| Bw.lbs2 | 18.5185 | 16.2346 | 20.6790 | 18.4774 | 11.0494 | 15.2469 | 11.1728 | 12.4897 | 11.17038 | 10.58626 | 11.55117 | 11.1026 |
| Cb.lbs1 | 35.4173 | 22.6482 | 37.7470 | 31.9375 | 13.2409 | 16.1014 | 12.8281 | 14.0568 | 18.28575 | 15.51547 | 18.79482 | 17.5320 |
| Cc.lbs1 | 15.3501 | 13.6894 | 20.4219 | 16.4871 | 15.5745 | 16.8761 | 13.8689 | 15.4398 | 14.21990 | 13.62440 | 15.15508 | 14.3331 |
| Co.lbs2 | 22.8309 | 21.2868 | 15.7353 | 19.9510 | 16.2868 | 14.9632 | 18.4559 | 16.5686 | 16.57206 | 16.28053 | 15.09326 | 15.9819 |

**Table 3C: Shallow LBSs**

| STRUCTURE | T.S. Method | | | | C.P. Method | | | | CM-to-CM Method | | | |
|---|---|---|---|---|---|---|---|---|---|---|---|---|
| | lig | res | sdc | ave | lig | res | sdc | ave | lig | res | sdc | ave |
| Ac.lbs1 | 67.5290 | 48.9961 | 51.3513 | 55.9588 | 0.6564 | 3.9382 | 3.2046 | 2.59973 | 22.61821 | 19.74564 | 20.04670 | 20.8035 |
| Ac.lbs2 | 81.9305 | 59.3050 | 62.7027 | 67.9794 | 0.1158 | 1.8147 | 1.3900 | 1.10683 | 24.89007 | 21.42218 | 21.93579 | 22.7493 |
| Ai.lbs2 | 86.8560 | 68.7221 | 76.5112 | 77.3631 | 0.0811 | 3.8540 | 1.7039 | 1.87967 | 25.36921 | 21.32868 | 22.85914 | 23.1857 |
| Ai.lbs3 | 91.7647 | 72.9817 | 74.0365 | 79.5943 | 0.0000 | 1.2170 | 0.8925 | 0.70317 | 27.08570 | 22.09977 | 22.36949 | 23.8517 |
| Al.lbs2 | 98.4760 | 92.8488 | 95.0762 | 95.4670 | 0.0000 | 1.6413 | 0.9379 | 0.85973 | 27.16972 | 24.73393 | 25.29898 | 25.7342 |
| At.lbs1 | 90.2678 | 82.8172 | 90.6403 | 87.9084 | 0.6286 | 1.9558 | 0.9546 | 1.17967 | 29.30139 | 27.18828 | 29.48152 | 28.6571 |
| At.lbs2 | 90.6403 | 59.1851 | 71.3620 | 73.7291 | 0.3027 | 4.4237 | 2.4680 | 2.39813 | 29.48389 | 23.00176 | 24.95588 | 25.8138 |
| At.lbs3 | 98.1141 | 88.6380 | 90.3143 | 92.3555 | 0.1164 | 1.3038 | 1.0477 | 0.82263 | 33.37043 | 28.74042 | 29.35302 | 30.4880 |
| At.lbs4 | 99.2317 | 94.9243 | 97.0664 | 97.0741 | 0.0698 | 1.2573 | 0.8847 | 0.73727 | 35.13666 | 31.21572 | 32.45906 | 32.9371 |
| Au.lbs3 | 96.9936 | 82.7487 | 90.4438 | 90.0620 | 0.3937 | 2.8275 | 1.7538 | 1.65833 | 31.85904 | 26.68224 | 28.65107 | 29.0641 |
| Av.lbs3 | 60.5753 | 69.7970 | 75.6345 | 68.6689 | 4.7377 | 2.0305 | 1.4382 | 2.73547 | 16.04120 | 17.21665 | 18.04765 | 17.1018 |
| Ba.lbs1 | 91.8503 | 82.4848 | 83.8716 | 86.0689 | 1.6337 | 3.7234 | 3.3815 | 2.91287 | 34.87654 | 31.58646 | 31.91738 | 32.7935 |
| Ba.lbs2 | 96.7325 | 90.7105 | 92.7622 | 93.4017 | 0.6459 | 2.0897 | 1.4628 | 1.39947 | 37.45268 | 34.36086 | 35.28394 | 35.6992 |
| Ba.lbs3 | 83.0737 | 82.6368 | 87.0251 | 84.2452 | 2.8875 | 3.5334 | 2.7166 | 3.04583 | 31.74831 | 31.62383 | 32.87885 | 32.0837 |
| Bb.lbs5 | 67.9236 | 43.9236 | 57.7580 | 56.5351 | 0.2038 | 1.9108 | 0.4586 | 0.85773 | 24.45640 | 20.55410 | 22.77531 | 22.5953 |
| Bc.lbs0 | 97.9282 | 82.1823 | 74.9309 | 85.0138 | 0.0000 | 0.4834 | 0.8287 | 0.43737 | 20.88671 | 18.12214 | 17.23462 | 18.7478 |
| Bg.lbs7 | 53.5948 | 51.5523 | 44.0632 | 49.7368 | 5.9913 | 6.1547 | 8.2244 | 6.79013 | 22.67510 | 22.27142 | 20.67289 | 21.8731 |
| Bm.lbs2 | 90.8924 | 77.0782 | 84.4743 | 84.1483 | 0.3667 | 2.0782 | 0.9780 | 1.14097 | 21.18420 | 19.04630 | 20.07942 | 20.1033 |
| Bn.lbs1 | 71.1596 | 54.6581 | 50.7929 | 58.8702 | 0.1982 | 3.7661 | 3.5679 | 2.51073 | 18.79190 | 17.07170 | 16.63080 | 17.4981 |
| Bo.lbs2 | 92.3749 | 64.6545 | 69.0230 | 75.3508 | 0.0794 | 3.4154 | 2.8594 | 2.11807 | 19.41146 | 15.52336 | 16.04409 | 16.9930 |
| Bo.lbs3 | 79.4281 | 65.6871 | 55.6791 | 66.9314 | 1.5091 | 3.8920 | 4.6863 | 3.36247 | 17.07288 | 15.62567 | 14.70304 | 15.8005 |
| Bp.lbs2 | 96.1934 | 63.6858 | 79.2145 | 79.6979 | 0.0000 | 3.2628 | 0.6042 | 1.28900 | 23.57125 | 17.58167 | 19.66017 | 20.2710 |
| Bp.lbs3 | 94.6224 | 60.7251 | 80.8459 | 78.7311 | 0.0000 | 3.6254 | 0.6042 | 1.40987 | 22.70205 | 17.16216 | 19.85317 | 19.9058 |
| Bp.lbs4 | 72.9909 | 51.6012 | 56.9184 | 60.5035 | 0.4834 | 4.4713 | 3.2628 | 2.73917 | 18.68024 | 16.06110 | 16.80683 | 17.1827 |
| Bq.lbs9 | 40.3425 | 45.0816 | 49.5818 | 45.0020 | 1.0753 | 0.7965 | 0.4779 | 0.78323 | 17.47268 | 18.26699 | 18.83646 | 18.1920 |
| Br.lbs4 | 84.9789 | 67.8903 | 69.8312 | 74.2335 | 0.1688 | 1.0549 | 1.0127 | 0.74547 | 24.67343 | 21.37890 | 21.69178 | 22.5814 |
| Br.lbs6 | 79.5781 | 63.9662 | 71.5612 | 71.7018 | 0.1266 | 1.2236 | 0.4641 | 0.60477 | 23.53866 | 20.67673 | 21.98487 | 22.0668 |
| Bt.lbs2 | 97.7396 | 89.8702 | 91.3771 | 92.9956 | 0.0419 | 1.0465 | 0.6279 | 0.57210 | 29.67364 | 26.54252 | 27.09555 | 27.7706 |
| Bt.lbs3 | 80.0335 | 62.2436 | 65.1319 | 69.1363 | 0.0837 | 1.5906 | 1.0046 | 0.89297 | 23.99113 | 20.65617 | 21.31436 | 21.9872 |
| Bt.lbs4 | 69.4851 | 45.1235 | 58.1833 | 57.5973 | 1.0465 | 6.9067 | 2.8882 | 3.61380 | 22.03106 | 18.02183 | 19.96236 | 20.0051 |
| Bu.lbs0 | 69.9014 | 54.7005 | 56.8613 | 60.4877 | 2.9189 | 7.9985 | 6.0273 | 5.64823 | 22.67629 | 20.07973 | 20.40709 | 21.0544 |
| Bv.lbs2 | 96.4315 | 87.1616 | 89.8277 | 91.1403 | 0.0000 | 1.2715 | 0.8203 | 0.69727 | 29.81212 | 25.88404 | 26.62986 | 27.4420 |
| Bw.lbs1 | 97.9012 | 70.1852 | 81.8519 | 83.3128 | 0.0000 | 3.3951 | 1.7901 | 1.72840 | 23.73275 | 18.22958 | 19.78687 | 20.5831 |
| Bz.lbs2 | 62.4858 | 68.9841 | 63.1385 | 64.8695 | 6.8956 | 4.2849 | 6.6402 | 5.94023 | 26.79193 | 28.44772 | 26.98328 | 27.4076 |
| Cc.lbs2 | 92.8187 | 70.4668 | 75.0898 | 79.4584 | 0.2693 | 3.4111 | 2.3339 | 2.00477 | 27.25275 | 22.81167 | 23.49499 | 24.5198 |
| Cc.lbs3 | 97.3519 | 74.9102 | 82.2262 | 84.8294 | 0.0000 | 2.5583 | 1.1670 | 1.24177 | 29.04908 | 23.44567 | 24.82153 | 25.7721 |
| Ck.lbs0 | 61.9676 | 67.1266 | 61.6077 | 63.5673 | 9.7181 | 8.3983 | 9.8980 | 9.33813 | 20.36178 | 21.22158 | 20.21158 | 20.5983 |
| Cn.lbs3 | 76.6895 | 67.8523 | 68.9740 | 71.1719 | 1.2859 | 3.5021 | 3.6662 | 2.81807 | 31.08082 | 28.14889 | 28.44036 | 29.2234 |